%% file: main.tex
\documentclass[%
 superscriptaddress,
 aip,
 amsmath,amssymb,
 reprint,%
longbibliography
]{revtex4-2}
\usepackage[utf8]{inputenc}
\usepackage[english]{babel}
\usepackage{color} 
\usepackage{graphicx} 
\usepackage{amsmath,amsthm,amsfonts,amssymb,amscd}
\usepackage{calc} 
\usepackage{array} 
\usepackage{titlesec} 
\usepackage{enumitem} 
\usepackage{changepage} 
\usepackage{hyperref} 
\usepackage{miller} 
\usepackage{soul} 
\usepackage{setspace} 
\usepackage{lipsum}
\usepackage{letltxmacro}  
\usepackage{float} 
\usepackage{mhchem}
\usepackage{textcomp}
\usepackage{capt-of}  
\usepackage{siunitx}
\usepackage{lineno}

\newcommand\mytitle{\textit{Operando} Electron Microscopy of Nanoscale Electronic Devices on Non-Conductive Substrates}

\input{header.tex}

\begin{document}

\title{\mytitle}

\author{Menglin Zhu}\thanks{These authors contributed equally to this work}
\affiliation{Department of Materials Science and Engineering, Massachusetts Institute of Technology, Cambridge, MA 02139}

\author{Michael Xu}\thanks{These authors contributed equally to this work}
\affiliation{Department of Materials Science and Engineering, Massachusetts Institute of Technology, Cambridge, MA 02139}

\author{Zishen Tian}\thanks{These authors contributed equally to this work}
\affiliation{Department of Materials Science and Engineering, University of California, Berkeley, Berkeley, CA 94720}
\affiliation{Rice Advanced Materials Institute, Rice University, Houston, TX 77005}

\author{Colin Gilgenbach}
\affiliation{Department of Materials Science and Engineering, Massachusetts Institute of Technology, Cambridge, MA 02139}

\author{Daniel Drury}
\affiliation{U.S. Army Combat Capabilities Development Command–Army Research Laboratory, Adelphi, Maryland 20783}

\author{Bridget R. Denzer}
\affiliation{Department of Materials Science and Engineering, Massachusetts Institute of Technology, Cambridge, MA 02139}

\author{Ching-Che Lin}
\affiliation{Applied Science and Technology Program, University of California, Berkeley, Berkeley, CA 94720}
\affiliation{Rice Advanced Materials Institute, Rice University, Houston, TX 77005}

\author{Deokyoung Kang}
\affiliation{Department of Materials Science and Engineering, University of California, Berkeley, Berkeley, CA 94720}
\affiliation{Rice Advanced Materials Institute, Rice University, Houston, TX 77005}

\author{Lane W. Martin}
\affiliation{Rice Advanced Materials Institute, Rice University, Houston, TX 77005}
\affiliation{Department of Materials Science and NanoEngineering, Rice University, Houston, TX 77005}
\affiliation{Department of Physics and Astronomy, Rice University, Houston, TX 77005}
\affiliation{Department of Chemistry, Rice University, Houston, TX 77005}

\author{James M.~LeBeau}
\email{lebeau@mit.edu}
\affiliation{Department of Materials Science and Engineering, Massachusetts Institute of Technology, Cambridge, MA 02139}

\date{\today}

\begin{abstract}
Achieving operating conditions comparable to ``bulk'' electronic devices, such as thin film capacitors, during \textit{operando} electron microscopy remains challenging, particularly when devices are grown on non-conductive substrates. Limited precision of focused ion beam milling for sample preparation often necessitates the use of conductive substrates or artificially thick layers that differ from actual device architectures. These modifications can alter native strain, electrostatic boundary conditions, and ultimately device response. Here, we present a generic and versatile workflow for \textit{operando} biasing of thin-film capacitors in the (scanning) transmission electron microscope, including sample fabrication and device operation. By introducing a patterned insulating barrier adjacent to the bulk-characterized capacitors, our approach enables sample preparation without altering the original film structure. As a case study, we apply the method to a piezoelectric thin-film capacitor grown on an insulating substrate, and demonstrate that it preserves the boundary-condition-sensitive domain switching at the atomic scale under applied electric fields. Overall, the process can help to establish a foundation for systematic \textit{operando} studies of complex thin-film systems under representative bulk testing geometries.

\end{abstract}

\maketitle

The response of materials to electric fields is central to the functionality of many electronic devices, for example, ferroelectrics, piezoelectrics, and ionic conductors. Synthesis of these materials in thin-film forms offers additional tunability through epitaxial strain \cite{Fernandez2022-bj, Schlom2007-la, Lee2014-zc, Lichtensteiger2023-eb, Nguyen2011-hr, Xu2020-hf, Kavle2022-tl, Kim2019-yh, Lee2014-zc}, interfacial engineering \cite{Hwang2012-vl, Shi2023-au, Nelson2011-zd, Yadav2016-ob, Abid2021-pe, Zhu2025-rm, Xie2017-zb}, and dimensional scaling \cite{Junquera2003-ip, Tenne2009-ka, Lee2015-au, Maksymovych2012-ul, Kim2005-sj, Chang2011-xe}. This tunability is inherently linked to the device structure \cite{Lichtensteiger2023-eb, Fernandez2022-bj}, film inhomogeneities \cite{Lee2014-zc, Kavle2022-tl, Zhu*2025-ub}, and boundary conditions \cite{Xie2017-zb, Schlom2007-la}. As a result, understanding and controlling such behavior requires spatially resolved, real-time studies capable of probing how local structural and chemical features mediate device functionality.

The application of an electric field within the (scanning) transmission electron microscope (STEM) enables \textit{in situ} and/or \textit{operando} characterization of structure and chemistry with nanometer-to-atomic-scale resolution. This approach has revealed a wide range of complex switching phenomena, including domain nucleation and domain-wall motion \cite{Li2019-en, Nelson2011-gk}, as well as field-induced phase transformations \cite{Pan*2024-xy, Nukala2021-tp}. It has also provided insight into defect dynamics, such as the critical role of oxygen migration in processes of delithiation in batteries and reversible phase transitions in hafnia-based ferroelectric devices \cite{Zhang2017-mo, Gong2018-ei, Nukala2021-tp, Recalde-Benitez2023-vd}. To directly connect these observations to functional behavior in actual devices, however, it is essential to preserve the film’s native electrostatic and mechanical boundary conditions during sample preparation for \textit{operando} STEM. 

This process is typically achieved through focused ion beam (FIB) milling and lift-out \cite{Recalde-Benitez2023-vd, Zhang2017-mo, Gong2018-ei, Pan*2024-xy, Nukala2021-tp, Recalde-Benitez2024-tq} onto micro-electromechanical systems (MEMS) chips with pre-patterned electrodes. The FIB-prepared cross-sectional sample is transferred using a nano-manipulator onto an electron-transparent window between the electrodes and mounted by metal deposition or sometimes electrostatic force to mitigate leakage current \cite{Recalde-Benitez2024-tq}. This configuration supports both bulk crystal biasing and in-plane biasing of thin films.

\begin{figure}[h]
    \centering
    \includegraphics[width=3.5in]{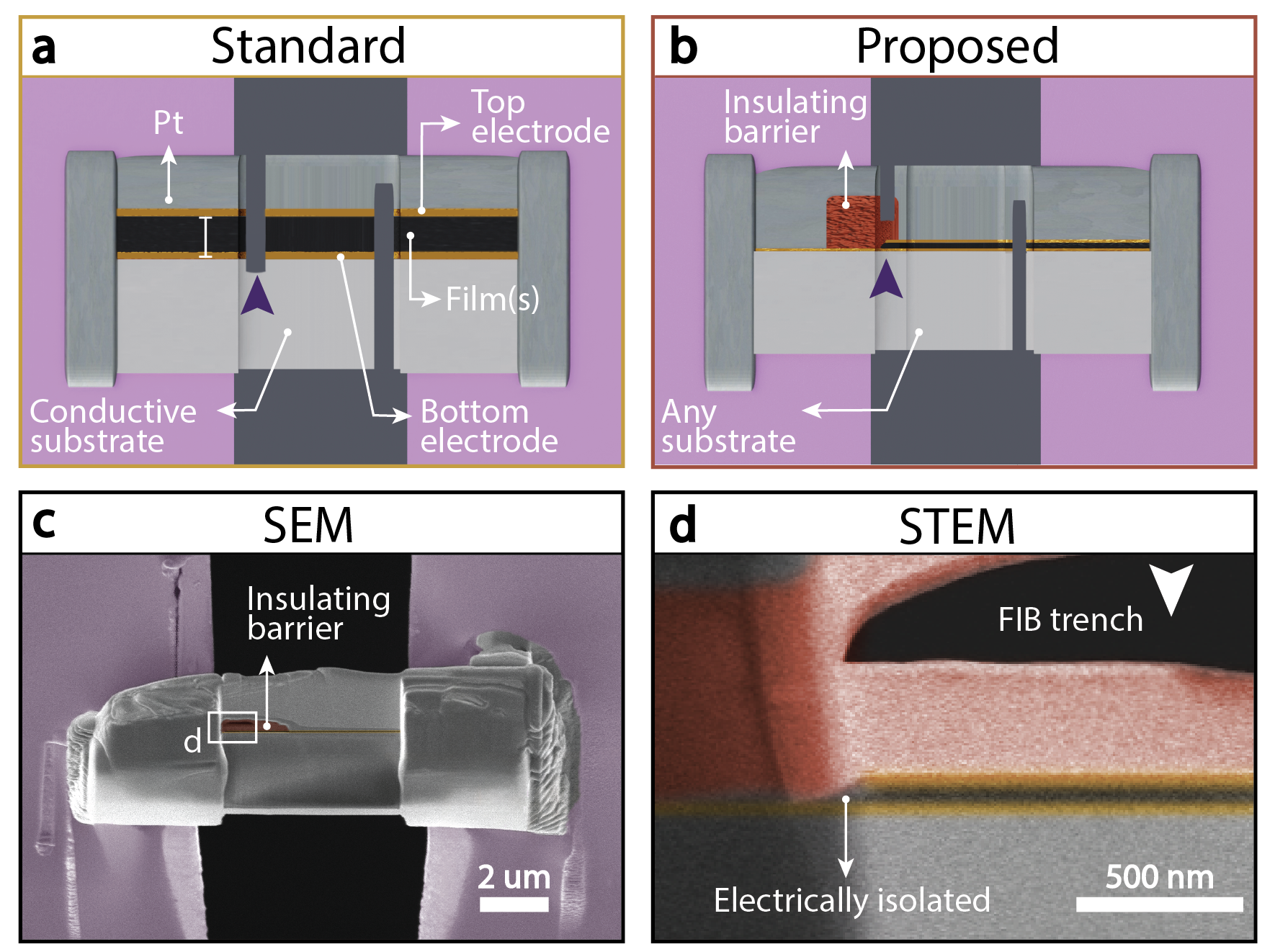}

    \caption{Schematic of (a) standard sample preparation, where FIB trenches terminate in the conductive substrate, or in the thin film or bottom electrode (within the capped line) if the substrate is insulating, and (b) the proposed method for arbitrarily thin samples, where trenches terminate in a thick insulating sacrificial layer, independent of film thickness or substrate type. (c) Experimental cross-sectional FIB image of a heterostructure comprising 25/80/25 nm of metal/dielectric/metal layers (false-colored), which are barely resolvable. (d) STEM cross-sectional image from the boxed region in (c), revealing the ultrathin layers and the cut made using the proposed method. Cuts are indicated with arrows.}
    \label{fig:intro}
\end{figure}

In contrast, applying an out-of-plane (OOP) bias (typical for bulk-capacitor operation) requires bottom and top electrodes to be isolated from one another \cite{Recalde-Benitez2023-vd, Zhang2017-mo, Recalde-Benitez2024-tq, Nukala2021-tp, Pan*2024-xy}. As shown in Figure~\ref{fig:intro}a, this is typically achieved by making two FIB cuts: one on the right, terminating in the FIB-deposited platinum to disconnect the bottom electrode, and another on the left, terminating in the film and/or bottom electrode to disconnect the top electrode. This setup, however, becomes increasingly impractical when the total capacitor thickness is below $\sim$100 nm. At such dimensions, the ion beam fails to spatially distinguish the layers (Figure~\ref{fig:intro}e-f), making precise electrical isolation of the electrodes extremely challenging (especially the left-hand cut). 

Alternatively, artificially thick layers or conductive substrates (\textit{e.g.}, Nb-doped \ce{SrTiO3} in place of un-doped\ce{SrTiO3}) are often employed  \cite{Han2014-rz, Pan*2024-xy, Nukala2021-tp, Zhang2017-mo, Recalde-Benitez2023-vd}. These modifications, however, can fundamentally change the electrostatic and mechanical boundary conditions. Increasing the layer thickness, for example, may also cause strain relaxation and/or affect the charge screening at the metal/dielectric (\textit{i.e.} thin films) interface, potentially modifying the behavior observed in macroscopic devices \cite{Morozovska2018-ed, Fernandez2022-bj, Lee2015-au, Tenne2009-ka, Maksymovych2012-ul, Junquera2003-ip}. Furthermore, conductive substrates that preserve the necessary strain conditions and band alignment, both of which are essential for achieving desired functional properties \cite{Schlom2007-la, Lee2014-zc, Fernandez2022-bj, Lichtensteiger2023-eb, Nguyen2011-hr, Xu2020-hf, Kavle2022-tl, Kim2019-yh}, are not always available (\textit{e.g.,} \textit{RE}ScO$_3$, \textit{RE} = Dy, Tb, Gd, Sm, Nd). 

In this Article, we propose a generic preparation method that is independent of the original thin-film architecture and substrate. By introducing a lithographically patterned non-conductive material around the rim of bulk capacitors, the most delicate milling step is shifted into the sacrificial layer (Figure~\ref{fig:intro}b, d), eliminating the need to directly mill into the thin layers for electrical isolation. We validate the proposed workflow on thin-film heterostructures with 80-nm-thick dielectric layers and 25-nm electrodes (with potential for arbitrarily thin layers) grown on insulating substrates. \textit{Operando} STEM measurements on the prepared samples demonstrate the efficacy of the approach by capturing domain evolution as a function of applied field magnitude and direction. This strategy provides a robust and generalizable framework for establishing metal/insulator/metal heterostructures suitable for \textit{operando} electron microscopy, while maintaining native device boundary conditions with bias applied out of out-of-plane.

\section{Materials and Methods}\label{sec:growth}

\subsection{Thin Film Growth and Capacitor Fabrication} 
Epitaxial heterostructures of 25\,nm \ce{Ba_{0.5}Sr_{0.5}RuO_{3}} / 30\,nm \ce{PbZr_{0.2}Ti_{0.8}O_{3}} / 20\,nm \ce{0.68PbMg_{1/3}Nb_{1/3}O_{3}-0.32PbTiO_{3}} / 30\,nm \ce{PbZr_{0.2}Ti_{0.8}O_{3}} / 25\,nm \ce{Ba_{0.5}Sr_{0.5}RuO_{3}} (BSRO/PZT/PMN-PT/PZT/BSRO) were deposited on \hkl(110)-oriented single-crystalline \ce{NdScO_{3}} substrates via pulsed-laser deposition (PLD). All depositions were performed using a KrF excimer laser ($\lambda$ = 248 nm, LPX-500, Coherent) in a vacuum chamber (Twente Solid State Technology) with an on-axis geometry and 5.5 cm target-to-substrate distance. The bottom BSRO layers were deposited at a heater temperature of 775 \unit{\degreeCelsius}, under a dynamic oxygen pressure of 100 mTorr, and at a laser fluence of 1.3 \si{J.cm^{-2}} and a laser repetition rate of 2 Hz. The PZT layers were deposited at a heater temperature of 600 \unit{\degreeCelsius}, in a dynamic oxygen pressure of 200 mTorr, and at a laser fluence of 1.7 \si{J.cm^{-2}} and a laser repetition rate of 10 Hz. The PMN-PT layers were deposited at the same conditions as the PZT layers except using a laser repetition rate of 5 Hz. The top BSRO layers were deposited \textit{in situ} following the growth of the dielectric layer(s) at the same conditions as the bottom metal layers, except keeping the heater temperature at 600 \unit{\degreeCelsius}, which minimizes the evaporation of volatile species (lead and magnesium). After deposition, the heterostructures were cooled to room temperature in a static oxygen pressure of 760 Torr with a 10 \si{\degreeCelsius.min^{-1}} cooling rate. 

\begin{figure*}
\begin{center}
    \includegraphics[width=1.0\textwidth]{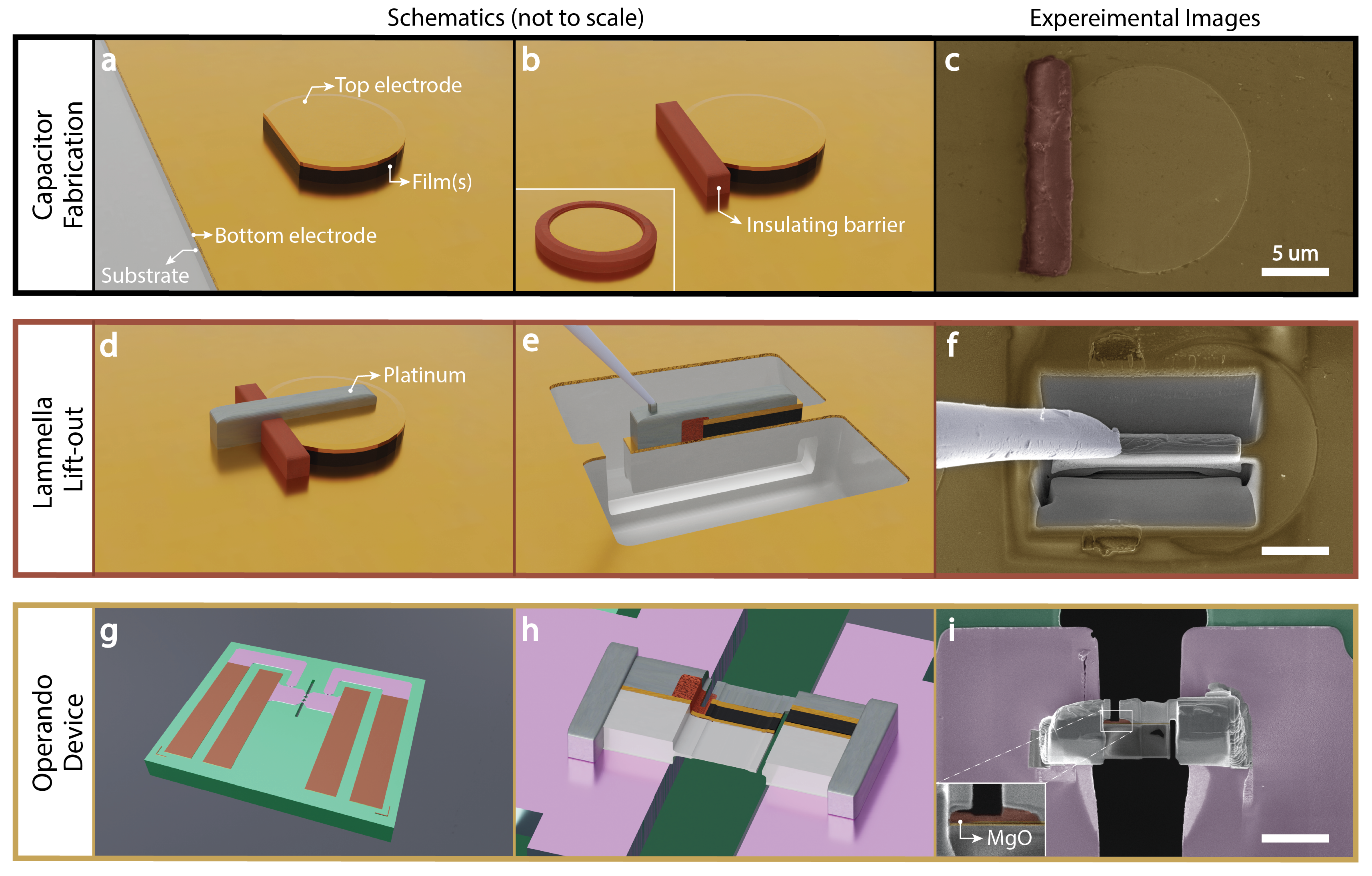}
    \caption{
        Schematics and corresponding false-colored SEM/FIB images of the sample preparation procedure.  
        (a) The top electrode and thin film(s) are selectively etched in a rectangular region at the edge of the circular capacitor, exposing the bottom electrode. The substrate on the left is shown only to illustrate the layered structure; the bottom electrodes remain intact.
        (b) A rectangular insulating layer (\ce{MgO} here) is deposited within the etched region, with the corresponding SEM image shown in (c).  
        (d) A \ce{Pt} protection layer is deposited across the \ce{MgO} insulating barrier.  
        (e) A cross-sectional lamella is lifted out using an Omniprobe following standard procedures of trenching and undercutting; the corresponding SEM image is shown in (f).
        (g) The lamella is transferred to a dedicated MEMS chip and (h) thinned to its final shape, with the SEM image shown in (i). The \ce{MgO} barrier enables reliable electrical isolation by trenching into $\sim$500 nm-thick insulating layer, which is well within the milling precision of modern FIB systems. Inset shows a magnified SEM image of the trench, corresponding to the same region imaged by STEM in Figure~\ref{fig:intro}d.
        The same scale bar in (c) applies to (f) and (i).
    }
    \label{fig:schematic}
\end{center}
\end{figure*}

Out-of-plane circular capacitors with diameters of 25 \si{\um} were were fabricated on the metal/dielectric/metal heterostructures via standard photolithography and ion milling techniques (Figure~\ref{fig:schematic}a). First, a photoresist pattern was made consisting of an array of circular capacitors. A G-line positive photoresist (OCG 825, Fujifilm) layer was spin-coated onto the surface, exposed through a shadow mask using a ultraviolet light source for 10 s, and developed in a tetramethylammonium-hydroxide developer (MF-319, Kayaku) diluted in deionized water (1:1) for 20-30 s. The region not covered by the photoresist pattern was then etched using an Ar$^{+}$ ion mill (Nanoquest Pico, Intlvac). The ion-milling process was precisely controlled to stop at the interface between the dielectric and bottom electrode, exposing the bottom electrode for electrical contact while leaving it intact (although, in principle, the method remains applicable provided the top electrodes are completely removed and the dielectric layer is preserved); end-point detection was enabled by \textit{in situ} secondary-ion mass spectrometry (SIMS; IMP-EPD, Hiden Analytical) which monitors the yield of various species (lead, zirconium, niobium, barium, strontium, and ruthenium) in real time. The photoresist pattern was removed afterwards via sonication in acetone for 1 minute. 
 
Next, a localized \ce{MgO} insulating barrier was fabricated immediately against the rim of each circular capacitor (Figure~\ref{fig:schematic}b), leveraging a second photolithography step and subsequent deposition. A fresh G-line positive photoresist (OCG 825) layer was spin-coated, exposed, and developed, defining a 3 $\times$ 50 \si{\um^{2}} rectangle tangent to and overlapping with the capacitor edge. The long axis of the rectangle was oriented roughly parallel to the desired STEM cross-sectional viewing direction. The \ce{MgO} layer was then deposited via room-temperature PLD, using a laser fluence of 3.0 \si{J.cm^{-2}}, repetition rate of 15 Hz, and a dynamic oxygen atmosphere of 20 mTorr. The nominal thickness of the \ce{MgO} layer is roughly 500 nm as to exceed the milling resolution of FIB. Alternatively, a circular insulating ring can be patterned around the capacitor stack (Figure~\ref{fig:schematic}b, inset), allowing extraction of the cross sectional lamella from any in-plane orientation instead of the fixed one dictated by the rectangular insulating barrier.

\subsection{\textit{Operando} Device FIB Preparation}

Cross-sectional STEM lamellae were prepared from the patterned capacitor following a standard FIB lamella lift-out procedure with a Thermo Fisher Scientific Helios Nanolab 600 DualBeam FIB/SEM. Scanning electron microscopy (SEM) overview images were acquired during sample preparation using the same instrument. To begin the FIB lift-out process, a 1.5 $\mu$m-wide, 3 $\mu$m-thick platinum protection strip was deposited orthogonal to and over the \ce{MgO} barrier, with a total length 2-4 $\mu$m longer than electrode gap on the MEMS chip (Figure~\ref{fig:schematic}g-i). Trenches were then milled on either side of this protected region to define the lamella, which was subsequently undercut and lifted out using an OmniProbe 400 nano-manipulator (Figure~\ref{fig:schematic}h-i).

After lift-out, the lamella was mounted on the electrodes across the electron-transparent window of a MEMS biasing chip and secured by platinum welds at both sides (Figure~\ref{fig:schematic}h-i). The lamella was then coarse-thinned to approximately 600 nm using a 30 kV Ga ion beam at 0.46 nA by iteratively polishing each side while tilting the stage $\pm$ 3\textdegree about the lamella mid-plane. 

Next, the stage was rotated 180\textdegree{} so that the lamella cross section was oriented near-perpendicular to the ion beam (ion beam view in Figure~\ref{fig:schematic}i) to mill two isolation trenches for the out-of-plane biasing geometry. The first $\sim1\mu$m-wide trench, adjacent to \ce{MgO} barrier, was cut from the top platinum protection layer and terminated in the \ce{MgO} barrier, thereby electrically isolating the top electrode and preserving the bottom electrode connection (Figure~\ref{fig:schematic}i, inset and Figure~\ref{fig:intro}d). The second trench on the opposite side cuts across the substrate and full film thickness, terminating in the top electrode (or platinum protection layer) and thus leaving it electrically connected. This configuration ensured that the primary electrical path through the lamella was along the out-of-plane direction only.

\subsection{Electron Microscopy and Analysis}

A Thermo Fisher Scientific Themis Z S/TEM was operated at 200~kV for STEM characterization. The probe semi-convergence angle was 19~mrad  for atomic-resolution imaging and 0.8~mrad for nano-beam electron diffraction (NBED). An Electron Microscope Pixel Array Detector (EMPAD)~\cite{Tate2016-kc} was used to collect two-dimensional diffraction patterns over a 2-dimensional scan grid at a step size of $\sim$1.4~nm and a dwell time of 1~$\mu$s, resulting in a 4D-STEM dataset. Cepstral analysis \cite{Padgett2020-au} was used to subsequently extract variations in film lattice parameter from the 4D-STEM dataset using the substrate as the reference. In the atomic resolution datasets, two-dimensional Gaussian distributions were used to fit and measure their displacements using a custom Python script. The fitted position of lead columns were subtracted from the average position of its four nearest neighboring zirconium/titanium atoms.

\textit{Operando} biasing experiments in the electron microscope were conducted using a DENSsolutions Lightning holder with dedicated MEMS chips. The electric field was applied by biasing the top and bottom electrodes of the FIB lamella using a Keithley 2450 Source Measurement Unit (SMU), controlled via the DENSsolutions Impulse software. The experiment was conducted in a two-probe configuration using a DC bias, while recording the current with the SMU.

To reduce electrical noise arising from ground loops (\textit{i.e.}, multiple ground pathways), the SMU, holder, and force-low were all grounded to the microscope case. Furthermore, without appropriate precautions when connecting the force-low terminal to ground, any difference in potential between the force-high side and force-low/ground (\textit{e.g.}~from static charge accumulation or mishandling) would result in a discharge through the MEMS chip and hence the S/TEM lamella, potentially destroying the sample. To prevent such damage, the force-high and -low terminals were initially shorted, \textit{i.e.}, the force-high terminal was also connected to the microscope ground. Once the SMU was powered on and initialized, the short was opened and the experiment was then conducted. After the experiment, the SMU output voltage was set to zero, the force-high/low terminals were again shorted, and the holder was safely disconnected without risk of discharge through the sample. 

\section{Results \& Discussion}

To validate the proposed sample preparation method for \textit{operando} S/TEM, a \ce{PbZr_{0.2}Ti_{0.8}O_3}-based heterostructure grown on an \textit{insulating} NdScO$_3$ (110) substrate is investigated (details in \nameref{sec:growth}). At this composition, PZT adopts a tetragonal structure similar to \ce{PbTiO3}, with polarization aligned along the elongated crystallographic \textit{c}-axis, as indicated in Figure~\ref{fig:groundState}a\cite{Kukhar2006-yd}.

This system offers an an excellent test case as its phase and domain structures are highly sensitive to composition and boundary conditions \cite{Kukhar2006-yd, Damodaran2017-kj, Lichtensteiger2023-eb, Kavle2022-tl, Lu2019-cb}. For example, to minimize electrostatic and elastic energies imposed by the specific substrate and film architectures, regions with aligned polarization and crystallographic orientation (ferroelectric/elastic domains) within the film can self-organize into distinct configurations \cite{Damodaran2017-kj}, transitioning from single-domain \textit{c} to polydomain \textit{c/a} and \textit{a$_1$/a$_2$} states, see Figure~\ref{fig:groundState}a \cite{Damodaran2017-kj, Lu2019-cb}. 

\subsection{Ground-State Domain Structure}\label{sec:exp_groundstate}

\begin{figure}[h]
    \centering
    \includegraphics[width=3.5in]{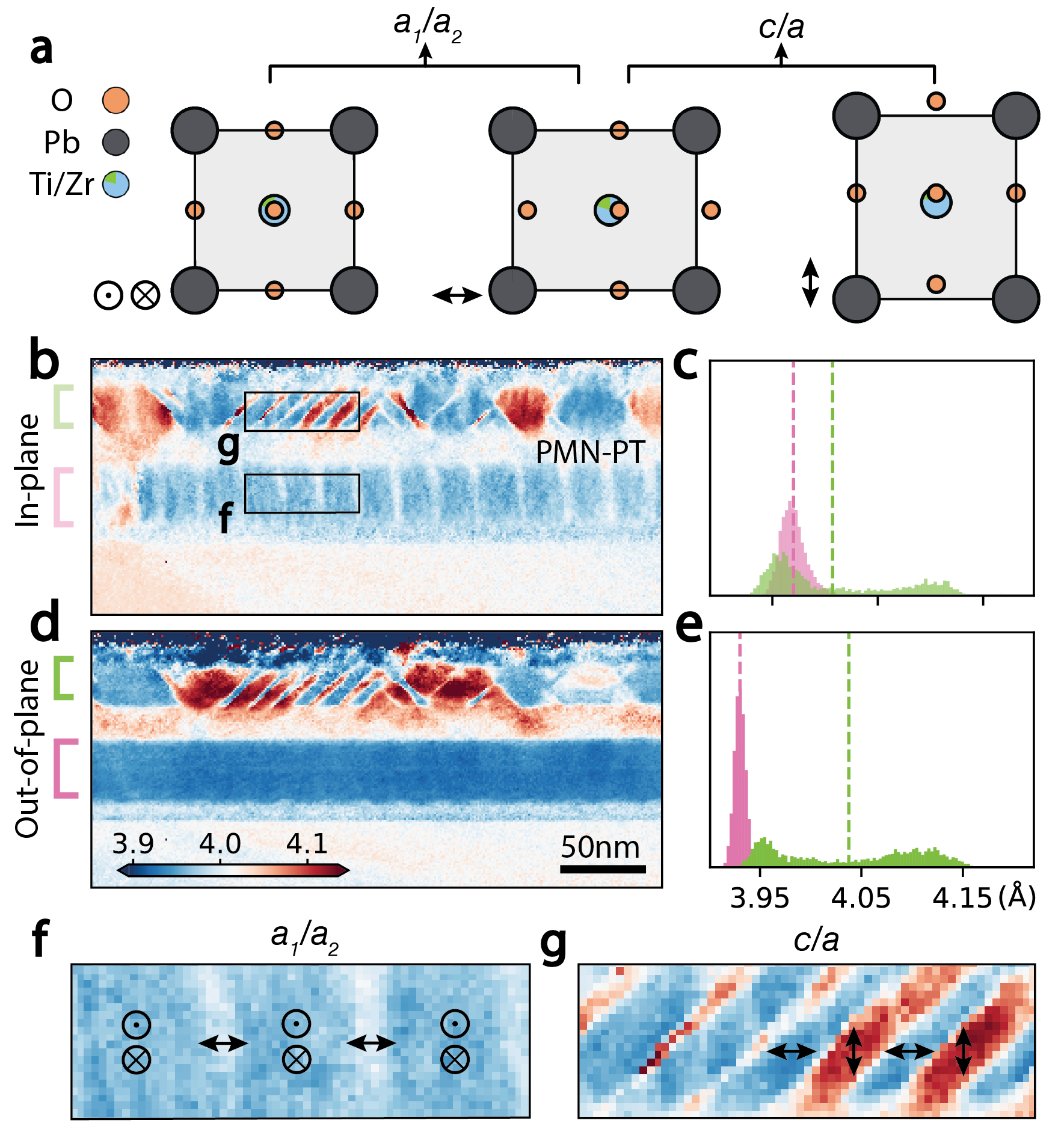}
    \caption{(a) Schematic models of each domain type, where polarization (arrows) is coupled to the elongated crystallographic axis. Cepstral-mapped (b) in-plane and (d) out-of-plane lattice parameters for BSRO/PZT/PMN-PT/PZT/BSRO \textit{operando} devices and (c, e) histograms for their corresponding distributions with mean indicated with dashed lines. Magnified regions from (b) highlight domain structures: (f) in-plane lattice fluctuations characteristic of the \textit{a$_1$/a$_2$} domain pattern, and (g) \textit{c}/\textit{a} domains, with polarization directions indicated by arrows.}
    \label{fig:groundState}
\end{figure}

A cross-sectional lamella is prepared from the PZT-based heterostructure following the steps stated above. Before the application of the electric field, NBED cepstral strain analysis \cite{Padgett2020-au} is applied to spatially map strain and domains in the ground state. While the tensile strain ($+0.3\%$) imposed by the NdScO$_3$ (110) substrate is expected to favor a mixed-phase domain structure of PZT (mixture of \textit{c/a} and \textit{a$_1$/a$_2$} domain configurations) \cite{Pandya2019-zb, Damodaran2017-kj}, a split behavior is observed between the two PZT layers. In the bottom 30~nm of PZT, an \textit{a$_1$/a$_2$} domain configuration is present (Figure~\ref{fig:groundState}b, f), as indicated by the unimodal distributions of in-plane and out-of-plane lattice constants (Figure~\ref{fig:groundState}c and e, pink). In contrast, the top 30~nm PZT layer above the PMN-PT interlayer adopts an \textit{c/a} domain configuration (Figure~\ref{fig:groundState}g), leading to a bimodal splitting of both the in-plane and out-of-plane lattice constants (Figure~\ref{fig:groundState}c and e, green). 

This transition occurs despite the largely preserved in-plane lattice constants of the PMN-PT layer relative to the NdScO$_3$ substrate (Figure~\ref{fig:groundState}b), \textit{i.e.}, similar nominal epitaxial strain states in the bottom and top PZT layers. The absence of a \textit{c/a} domain configuration in the bottom PZT layer can be attributed to stronger clamping from the NdScO$_3$ substrate, which favors formation of \textit{a$_1$/a$_2$} domains under tensile strain to minimize elastic energy \cite{Pompe1993-dp}. In contrast, the \textit{c/a} configuration in the top PZT layer and its higher associated elastic energy is mitigated by the presence of the mechanically softer PMN-PT middle layer, which can adjust its polar state to accommodate adjacent ferroelectric layers, as reflected in the slightly larger mean in-plane lattice constant of the top compared to the bottom PZT layer (Figure~\ref{fig:groundState}c, dashed lines). The distinct domain configurations between the two layers underscore the strong sensitivity of polar domains to structural and electrostatic boundary conditions, reinforcing the need to preserve the original bulk device architecture for \textit{operando} S/TEM biasing.

\subsection{Domain Transformation}

\begin{figure}[h!]
    \centering
    \includegraphics[width=3.5in]{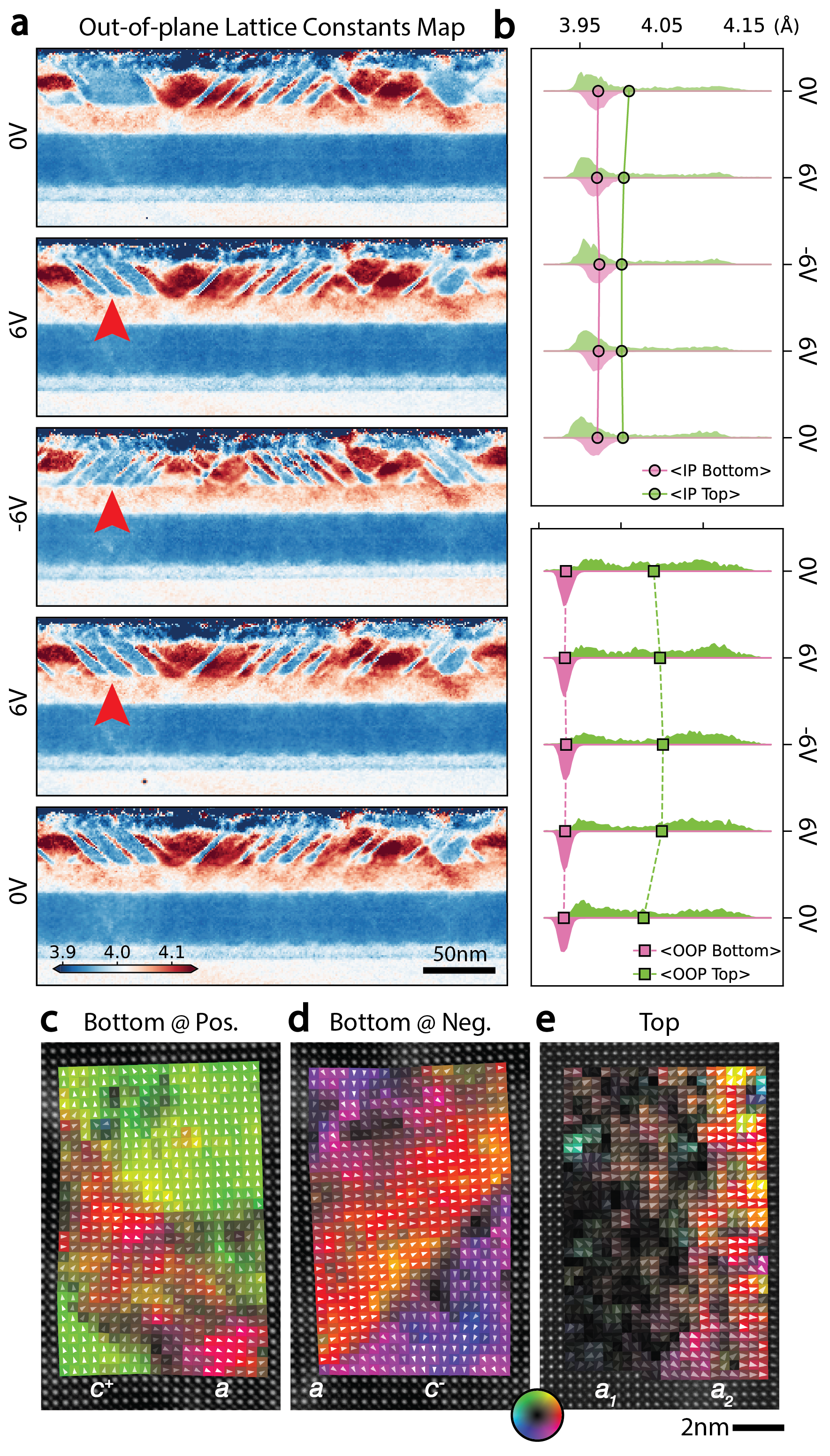}
    \caption{(a) Cepstral mapped out-of-plane lattice parameters of the same region from the \textit{operando} device throughout a full bias cycle. Field-induced and reoriented \textit{c} domains are marked with arrows. (b) Corresponding density distributions of in-plane and out-of-plane lattice parameters for the top and bottom PZT layers, with their respective mean values traced by lines. \textit{Operando} atomic-resolution images acquired from the identical region of the bottom layer under (c) positive and (d) negative bias, and (e) from the top layer. Displacements of Pb relative to neighboring Ti/Zr atoms are extracted from the image and overlaid as arrows. Arrow hue indicates displacement direction, while saturation reflects magnitude, ranging from 0 to 30 pm as shown in the color wheel.}
    \label{fig:change}
\end{figure}

While electrically biasing the PZT-based device through 0\,$\rightarrow$\,+6\,$\rightarrow$\,–6\,$\rightarrow$\,+6\,$\rightarrow$\,0\,V cycles, the domain configuration in the same sample region is monitored using NBED (Figure~\ref{fig:change}a-b) and atomic resolution high angle annular dark field (HAADF) STEM imaging (Figure~\ref{fig:change}c-e). When increasing the bias from 0 to 6 V, the film responds through the formation of additional \textit{c} domains within the \textit{a} domains in the top layer (indicated with red arrows in Figure~\ref{fig:change}a). This is a consequence of the lowered electrostatic energy prevailing over the increased elastic energy of this configuration. Upon field reversal (6 to –6 V), the orientation or polarization direction of the \textit{c} domains flips. Due to the coupling between ferroelectric and ferroelastic axes, the original interlaced \textit{c}/\textit{a} domain walls rotate from \hkl(101) to \hkl(-101) \cite{Ehara2017-cs, Zhang2019-dl}. 

The structural reorientation, driven by polarization-lattice coupling, is further visualized by \textit{operando} atomic resolution HAADF STEM images from the same region before and after switching (Figure~\ref{fig:change}c vs. d). Within the field of view, the domain boundaries and polar displacements within the domains flipped accordingly to accommodate the switch in field, highlighting the responsiveness of the heterostructure to bias even at the atomic scale. Such directional nature of this switching confirms that the observation is not a result of Joule heating from leakage, but instead reflects the fundamental mechanisms of switching, captured here on this non-conductive substrate that would otherwise be challenging to investigate.

In contrast to the active top layer, the bottom PZT layer beneath the PMN-PT insertion remains idle, retaining its \textit{a$_1$}/\textit{a$_2$} configuration throughout the entire voltage cycle, as confirmed by NBED and HAADF-STEM (Figure~\ref{fig:change}a,e). This spatially heterogeneous response across different layers underscores the value of \textit{operando} S/TEM imaging to reveal local phenomena that would otherwise be obscured by spatial averaging.

As one of the primary motivations for engineering domain structures in PZT is to harness the piezoelectric effect, \textit{i.e.}, the change in mechanical strain in response to an external electric field, tracking the average lattice parameters is critically important. For the bottom and top PZT layers (Figure~\ref{fig:change}b, lines), a clear difference in response is observed. The bottom layer maintains nearly constant lattice parameters in both in-plane and  out-of-plane directions, consistent with its stable \textit{a$_1$}/\textit{a$_2$} domain structure. In contrast, the top PZT layer exhibits a field-dependent increase in the average out-of-plane lattice parameter during field application, which reversibly returns to its original state once the field is removed. This strain modulation is attributed to the formation and annihilation of field-induced \textit{c} domains due to the coupling between ferroelectric and ferroelastic axes \cite{Damodaran2017-kj}. 

Together, these observations not only validate the efficacy of the proposed sample preparation process, but also demonstrate its broader potential for extending \textit{operando} S/TEM investigations to more complex heterostructures and layered systems where electrical, mechanical, or chemical constraints may simultaneously affect functional response.

\section{Conclusions}
Robust \textit{operando} S/TEM biasing of nanoscale electronic devices, such as thin film capacitors, can be achieved under bulk-like operating conditions, including the use of non-conductive substrates and arbitrarily thin layers. By incorporating a lithographically patterned insulating barrier at the rim of a capacitor, precise lamella extraction directly from bulk-characterized devices is enabled. This approach overcomes the limitations of conductive substrates or artificially thick layers that may change the boundary conditions, and further the performance, of the devices. As a test case, electric-field-induced domain switching in a PZT-based heterostructure is revealed, with spatially heterogeneous responses that highlight the strong sensitivity of ferroelectric domains to local mechanical and electrostatic constraints. These results underscore the importance of preserving bulk device structure to probe behavior in functional oxides, validating our proposed method as a reliable process for enabling systematic \textit{operando} investigations. The approach is broadly applicable to other ferroic, correlated, or electrochemically active thin-film systems, particularly those sensitive to substrate-induced boundary conditions.

\bibliography{paperpile.bib}

\section*{Acknowledgments}
This research was sponsored by the Army Research Laboratory and was accomplished under Cooperative Agreement Number W911NF-24-2-0100. The views and conclusions contained in this document and those of the authors should not be interpreted as representing the official policies, either expressed or implied, of the Army Research Laboratory or the U.S. Government. The U.S. Government is authorized to reproduce and distribute reprints for Government purposes, notwithstanding any copyright notation herein. This material is also based upon work supported by the Air Force Office of Scientific Research under award number FA9550-24-1-0266. Any opinions, findings, and conclusions or recommendations expressed in this material are those of the author(s) and do not necessarily reflect the views of the United States Air Force. This work made use of the MIT.nano Characterization Facilities. 

\section*{Author Information}
\subsection*{Contributions}
M.Z., M.X., and Z.T. proposed the design, and contributed equally to this work. M.Z., M.X., C.G., and B.R.D. performed sample preparation, electron microscopy, and data analysis. Z.T., D.D., C.C.L., and D.K. synthesized the heterostructures. J.M.L. and L.W.M. supervised the research. All authors co-wrote and edited the manuscript.

\section*{Ethics declarations}
\subsection*{Competing Interests}
The authors declare no competing interests.

\section*{Data Availability} 
Data supporting the manuscript and codes for image analysis are available from the corresponding authors by reasonable request.
\end{document}

%% file: header.tex
\definecolor{linkColor}{rgb}{0.7,0,0}
\definecolor{darkred}{rgb}{0.7,0,0}

\hypersetup{
    pdftitle={\mytitle},
    pdfauthor={Menglin Zhu},
    pdfpagemode=FullScreen,
    pdfborder={0 0 0},
    colorlinks=true,
    urlcolor=linkColor,
    citecolor=linkColor
}

\titleformat{\section}
{\color{darkred}\sffamily\bfseries}
{\color{darkred}\thesection}{1em}{}
\titleformat{\subsection}
{\color{darkred}\sffamily\itshape}
{\color{darkred}\thesubsection}{1em}{}
\titlespacing*{\section}
{0pt}{8pt}{0pt}
\titlespacing*{\subsection}
{0pt}{8pt}{0pt}

\emergencystretch=\maxdimen
\LetLtxMacro{\ORIGselectlanguage}{\selectlanguage}
\makeatletter
\DeclareRobustCommand{\selectlanguage}[1]{%
  \@ifundefined{alias@\string#1}
    {\ORIGselectlanguage{#1}}
    {\begingroup\edef\x{\endgroup
       \noexpand\ORIGselectlanguage{\@nameuse{alias@#1}}}\x}%
}
\newcommand{\definelanguagealias}[2]{%
  \@namedef{alias@#1}{#2}%
}
\makeatother

\definelanguagealias{en}{english}
\definelanguagealias{EN}{english}
\definelanguagealias{de}{english}

%% file: main.bbl
\begin{thebibliography}{41}%
\makeatletter
\providecommand \@ifxundefined [1]{%
 \@ifx{#1\undefined}
}%
\providecommand \@ifnum [1]{%
 \ifnum #1\expandafter \@firstoftwo
 \else \expandafter \@secondoftwo
 \fi
}%
\providecommand \@ifx [1]{%
 \ifx #1\expandafter \@firstoftwo
 \else \expandafter \@secondoftwo
 \fi
}%
\providecommand \natexlab [1]{#1}%
\providecommand \enquote  [1]{``#1''}%
\providecommand \bibnamefont  [1]{#1}%
\providecommand \bibfnamefont [1]{#1}%
\providecommand \citenamefont [1]{#1}%
\providecommand \href@noop [0]{\@secondoftwo}%
\providecommand \href [0]{\begingroup \@sanitize@url \@href}%
\providecommand \@href[1]{\@@startlink{#1}\@@href}%
\providecommand \@@href[1]{\endgroup#1\@@endlink}%
\providecommand \@sanitize@url [0]{\catcode `\\12\catcode `\$12\catcode `\&12\catcode `\#12\catcode `\^12\catcode `\_12\catcode `\%12\relax}%
\providecommand \@@startlink[1]{}%
\providecommand \@@endlink[0]{}%
\providecommand \url  [0]{\begingroup\@sanitize@url \@url }%
\providecommand \@url [1]{\endgroup\@href {#1}{\urlprefix }}%
\providecommand \urlprefix  [0]{URL }%
\providecommand \Eprint [0]{\href }%
\providecommand \doibase [0]{https://doi.org/}%
\providecommand \selectlanguage [0]{\@gobble}%
\providecommand \bibinfo  [0]{\@secondoftwo}%
\providecommand \bibfield  [0]{\@secondoftwo}%
\providecommand \translation [1]{[#1]}%
\providecommand \BibitemOpen [0]{}%
\providecommand \bibitemStop [0]{}%
\providecommand \bibitemNoStop [0]{.\EOS\space}%
\providecommand \EOS [0]{\spacefactor3000\relax}%
\providecommand \BibitemShut  [1]{\csname bibitem#1\endcsname}%
\let\auto@bib@innerbib\@empty
\bibitem [{\citenamefont {Fernandez}\ \emph {et~al.}(2022)\citenamefont {Fernandez}, \citenamefont {Acharya}, \citenamefont {Lee}, \citenamefont {Schimpf}, \citenamefont {Jiang}, \citenamefont {Lou}, \citenamefont {Tian},\ and\ \citenamefont {Martin}}]{Fernandez2022-bj}%
  \BibitemOpen
  \bibfield  {author} {\bibinfo {author} {\bibfnamefont {A.}~\bibnamefont {Fernandez}}, \bibinfo {author} {\bibfnamefont {M.}~\bibnamefont {Acharya}}, \bibinfo {author} {\bibfnamefont {H.-G.}\ \bibnamefont {Lee}}, \bibinfo {author} {\bibfnamefont {J.}~\bibnamefont {Schimpf}}, \bibinfo {author} {\bibfnamefont {Y.}~\bibnamefont {Jiang}}, \bibinfo {author} {\bibfnamefont {D.}~\bibnamefont {Lou}}, \bibinfo {author} {\bibfnamefont {Z.}~\bibnamefont {Tian}},\ and\ \bibinfo {author} {\bibfnamefont {L.~W.}\ \bibnamefont {Martin}},\ }\bibfield  {title} {\enquote {\bibinfo {title} {{Thin-film ferroelectrics}},}\ }\href {https://doi.org/10.1002/adma.202108841} {\bibfield  {journal} {\bibinfo  {journal} {Adv. Mater.}\ } (\bibinfo {year} {2022}),\ 10.1002/adma.202108841}\BibitemShut {NoStop}%
\bibitem [{\citenamefont {Schlom}\ \emph {et~al.}(2007)\citenamefont {Schlom}, \citenamefont {Chen}, \citenamefont {Eom}, \citenamefont {Rabe}, \citenamefont {Streiffer},\ and\ \citenamefont {Triscone}}]{Schlom2007-la}%
  \BibitemOpen
  \bibfield  {author} {\bibinfo {author} {\bibfnamefont {D.~G.}\ \bibnamefont {Schlom}}, \bibinfo {author} {\bibfnamefont {L.-Q.}\ \bibnamefont {Chen}}, \bibinfo {author} {\bibfnamefont {C.-B.}\ \bibnamefont {Eom}}, \bibinfo {author} {\bibfnamefont {K.~M.}\ \bibnamefont {Rabe}}, \bibinfo {author} {\bibfnamefont {S.~K.}\ \bibnamefont {Streiffer}},\ and\ \bibinfo {author} {\bibfnamefont {J.-M.}\ \bibnamefont {Triscone}},\ }\bibfield  {title} {\enquote {\bibinfo {title} {{Strain tuning of ferroelectric thin films}},}\ }\href {https://doi.org/10.1146/annurev.matsci.37.061206.113016} {\bibfield  {journal} {\bibinfo  {journal} {Annu. Rev. Mater. Res.}\ } (\bibinfo {year} {2007}),\ 10.1146/annurev.matsci.37.061206.113016}\BibitemShut {NoStop}%
\bibitem [{\citenamefont {Lee}\ \emph {et~al.}(2014)\citenamefont {Lee}, \citenamefont {Jeon}, \citenamefont {Yoon}, \citenamefont {Shin}, \citenamefont {Lee}, \citenamefont {Song}, \citenamefont {Bu}, \citenamefont {Kim}, \citenamefont {Chung}, \citenamefont {Yoon},\ and\ \citenamefont {Noh}}]{Lee2014-zc}%
  \BibitemOpen
  \bibfield  {author} {\bibinfo {author} {\bibfnamefont {D.}~\bibnamefont {Lee}}, \bibinfo {author} {\bibfnamefont {B.~C.}\ \bibnamefont {Jeon}}, \bibinfo {author} {\bibfnamefont {A.}~\bibnamefont {Yoon}}, \bibinfo {author} {\bibfnamefont {Y.~J.}\ \bibnamefont {Shin}}, \bibinfo {author} {\bibfnamefont {M.~H.}\ \bibnamefont {Lee}}, \bibinfo {author} {\bibfnamefont {T.~K.}\ \bibnamefont {Song}}, \bibinfo {author} {\bibfnamefont {S.~D.}\ \bibnamefont {Bu}}, \bibinfo {author} {\bibfnamefont {M.}~\bibnamefont {Kim}}, \bibinfo {author} {\bibfnamefont {J.-S.}\ \bibnamefont {Chung}}, \bibinfo {author} {\bibfnamefont {J.-G.}\ \bibnamefont {Yoon}},\ and\ \bibinfo {author} {\bibfnamefont {T.~W.}\ \bibnamefont {Noh}},\ }\bibfield  {title} {\enquote {\bibinfo {title} {{Flexoelectric control of defect formation in ferroelectric epitaxial thin films}},}\ }\href {https://doi.org/10.1002/adma.201400654} {\bibfield  {journal} {\bibinfo  {journal} {Adv. Mater.}\ } (\bibinfo {year} {2014}),\ 10.1002/adma.201400654}\BibitemShut
  {NoStop}%
\bibitem [{\citenamefont {Lichtensteiger}\ \emph {et~al.}(2023)\citenamefont {Lichtensteiger}, \citenamefont {Hadjimichael}, \citenamefont {Zatterin}, \citenamefont {Su}, \citenamefont {Gaponenko}, \citenamefont {Tovaglieri}, \citenamefont {Paruch}, \citenamefont {Gloter},\ and\ \citenamefont {Triscone}}]{Lichtensteiger2023-eb}%
  \BibitemOpen
  \bibfield  {author} {\bibinfo {author} {\bibfnamefont {C.}~\bibnamefont {Lichtensteiger}}, \bibinfo {author} {\bibfnamefont {M.}~\bibnamefont {Hadjimichael}}, \bibinfo {author} {\bibfnamefont {E.}~\bibnamefont {Zatterin}}, \bibinfo {author} {\bibfnamefont {C.-P.}\ \bibnamefont {Su}}, \bibinfo {author} {\bibfnamefont {I.}~\bibnamefont {Gaponenko}}, \bibinfo {author} {\bibfnamefont {L.}~\bibnamefont {Tovaglieri}}, \bibinfo {author} {\bibfnamefont {P.}~\bibnamefont {Paruch}}, \bibinfo {author} {\bibfnamefont {A.}~\bibnamefont {Gloter}},\ and\ \bibinfo {author} {\bibfnamefont {J.-M.}\ \bibnamefont {Triscone}},\ }\bibfield  {title} {\enquote {\bibinfo {title} {{Mapping the complex evolution of ferroelastic/ferroelectric domain patterns in epitaxially strained PbTiO3 heterostructures}},}\ }\href {https://doi.org/10.1063/5.0154161} {\bibfield  {journal} {\bibinfo  {journal} {APL Mater.}\ } (\bibinfo {year} {2023}),\ 10.1063/5.0154161}\BibitemShut {NoStop}%
\bibitem [{\citenamefont {Nguyen}\ \emph {et~al.}(2011)\citenamefont {Nguyen}, \citenamefont {Dekkers}, \citenamefont {Houwman}, \citenamefont {Steenwelle}, \citenamefont {Wan}, \citenamefont {Roelofs}, \citenamefont {Schmitz-Kempen},\ and\ \citenamefont {Rijnders}}]{Nguyen2011-hr}%
  \BibitemOpen
  \bibfield  {author} {\bibinfo {author} {\bibfnamefont {M.~D.}\ \bibnamefont {Nguyen}}, \bibinfo {author} {\bibfnamefont {M.}~\bibnamefont {Dekkers}}, \bibinfo {author} {\bibfnamefont {E.}~\bibnamefont {Houwman}}, \bibinfo {author} {\bibfnamefont {R.}~\bibnamefont {Steenwelle}}, \bibinfo {author} {\bibfnamefont {X.}~\bibnamefont {Wan}}, \bibinfo {author} {\bibfnamefont {A.}~\bibnamefont {Roelofs}}, \bibinfo {author} {\bibfnamefont {T.}~\bibnamefont {Schmitz-Kempen}},\ and\ \bibinfo {author} {\bibfnamefont {G.}~\bibnamefont {Rijnders}},\ }\bibfield  {title} {\enquote {\bibinfo {title} {{Misfit strain dependence of ferroelectric and piezoelectric properties of clamped (001) epitaxial Pb(Zr0.52,Ti0.48)O3 thin films}},}\ }\href {https://doi.org/10.1063/1.3669527} {\bibfield  {journal} {\bibinfo  {journal} {Appl. Phys. Lett.}\ } (\bibinfo {year} {2011}),\ 10.1063/1.3669527}\BibitemShut {NoStop}%
\bibitem [{\citenamefont {Xu}\ \emph {et~al.}(2020)\citenamefont {Xu}, \citenamefont {Huang}, \citenamefont {Barnard}, \citenamefont {Hong}, \citenamefont {Singh}, \citenamefont {Wong}, \citenamefont {Jansen}, \citenamefont {Harbola}, \citenamefont {Xiao}, \citenamefont {Wang}, \citenamefont {Crossley}, \citenamefont {Lu}, \citenamefont {Liu},\ and\ \citenamefont {Hwang}}]{Xu2020-hf}%
  \BibitemOpen
  \bibfield  {author} {\bibinfo {author} {\bibfnamefont {R.}~\bibnamefont {Xu}}, \bibinfo {author} {\bibfnamefont {J.}~\bibnamefont {Huang}}, \bibinfo {author} {\bibfnamefont {E.~S.}\ \bibnamefont {Barnard}}, \bibinfo {author} {\bibfnamefont {S.~S.}\ \bibnamefont {Hong}}, \bibinfo {author} {\bibfnamefont {P.}~\bibnamefont {Singh}}, \bibinfo {author} {\bibfnamefont {E.~K.}\ \bibnamefont {Wong}}, \bibinfo {author} {\bibfnamefont {T.}~\bibnamefont {Jansen}}, \bibinfo {author} {\bibfnamefont {V.}~\bibnamefont {Harbola}}, \bibinfo {author} {\bibfnamefont {J.}~\bibnamefont {Xiao}}, \bibinfo {author} {\bibfnamefont {B.~Y.}\ \bibnamefont {Wang}}, \bibinfo {author} {\bibfnamefont {S.}~\bibnamefont {Crossley}}, \bibinfo {author} {\bibfnamefont {D.}~\bibnamefont {Lu}}, \bibinfo {author} {\bibfnamefont {S.}~\bibnamefont {Liu}},\ and\ \bibinfo {author} {\bibfnamefont {H.~Y.}\ \bibnamefont {Hwang}},\ }\bibfield  {title} {\enquote {\bibinfo {title} {{Strain-induced room-temperature ferroelectricity in SrTiO3 membranes}},}\
  }\href {https://doi.org/10.1038/s41467-020-16912-3} {\bibfield  {journal} {\bibinfo  {journal} {Nat. Commun.}\ } (\bibinfo {year} {2020}),\ 10.1038/s41467-020-16912-3}\BibitemShut {NoStop}%
\bibitem [{\citenamefont {Kavle}\ \emph {et~al.}(2022)\citenamefont {Kavle}, \citenamefont {Zorn}, \citenamefont {Dasgupta}, \citenamefont {Wang}, \citenamefont {Ramesh}, \citenamefont {Chen},\ and\ \citenamefont {Martin}}]{Kavle2022-tl}%
  \BibitemOpen
  \bibfield  {author} {\bibinfo {author} {\bibfnamefont {P.}~\bibnamefont {Kavle}}, \bibinfo {author} {\bibfnamefont {J.~A.}\ \bibnamefont {Zorn}}, \bibinfo {author} {\bibfnamefont {A.}~\bibnamefont {Dasgupta}}, \bibinfo {author} {\bibfnamefont {B.}~\bibnamefont {Wang}}, \bibinfo {author} {\bibfnamefont {M.}~\bibnamefont {Ramesh}}, \bibinfo {author} {\bibfnamefont {L.-Q.}\ \bibnamefont {Chen}},\ and\ \bibinfo {author} {\bibfnamefont {L.~W.}\ \bibnamefont {Martin}},\ }\bibfield  {title} {\enquote {\bibinfo {title} {{Strain-driven mixed-phase domain architectures and topological transitions in Pb1- x Srx TiO3 thin films}},}\ }\href {https://doi.org/10.1002/adma.202203469} {\bibfield  {journal} {\bibinfo  {journal} {Adv. Mater.}\ } (\bibinfo {year} {2022}),\ 10.1002/adma.202203469}\BibitemShut {NoStop}%
\bibitem [{\citenamefont {Kim}\ \emph {et~al.}(2019)\citenamefont {Kim}, \citenamefont {Takenaka}, \citenamefont {Qi}, \citenamefont {Damodaran}, \citenamefont {Fernandez}, \citenamefont {Gao}, \citenamefont {McCarter}, \citenamefont {Saremi}, \citenamefont {Chung}, \citenamefont {Rappe},\ and\ \citenamefont {Martin}}]{Kim2019-yh}%
  \BibitemOpen
  \bibfield  {author} {\bibinfo {author} {\bibfnamefont {J.}~\bibnamefont {Kim}}, \bibinfo {author} {\bibfnamefont {H.}~\bibnamefont {Takenaka}}, \bibinfo {author} {\bibfnamefont {Y.}~\bibnamefont {Qi}}, \bibinfo {author} {\bibfnamefont {A.~R.}\ \bibnamefont {Damodaran}}, \bibinfo {author} {\bibfnamefont {A.}~\bibnamefont {Fernandez}}, \bibinfo {author} {\bibfnamefont {R.}~\bibnamefont {Gao}}, \bibinfo {author} {\bibfnamefont {M.~R.}\ \bibnamefont {McCarter}}, \bibinfo {author} {\bibfnamefont {S.}~\bibnamefont {Saremi}}, \bibinfo {author} {\bibfnamefont {L.}~\bibnamefont {Chung}}, \bibinfo {author} {\bibfnamefont {A.~M.}\ \bibnamefont {Rappe}},\ and\ \bibinfo {author} {\bibfnamefont {L.~W.}\ \bibnamefont {Martin}},\ }\bibfield  {title} {\enquote {\bibinfo {title} {{Epitaxial strain control of relaxor ferroelectric phase evolution}},}\ }\href {https://doi.org/10.1002/adma.201901060} {\bibfield  {journal} {\bibinfo  {journal} {Adv. Mater.}\ } (\bibinfo {year} {2019}),\ 10.1002/adma.201901060}\BibitemShut
  {NoStop}%
\bibitem [{\citenamefont {Hwang}\ \emph {et~al.}(2012)\citenamefont {Hwang}, \citenamefont {Iwasa}, \citenamefont {Kawasaki}, \citenamefont {Keimer}, \citenamefont {Nagaosa},\ and\ \citenamefont {Tokura}}]{Hwang2012-vl}%
  \BibitemOpen
  \bibfield  {author} {\bibinfo {author} {\bibfnamefont {H.~Y.}\ \bibnamefont {Hwang}}, \bibinfo {author} {\bibfnamefont {Y.}~\bibnamefont {Iwasa}}, \bibinfo {author} {\bibfnamefont {M.}~\bibnamefont {Kawasaki}}, \bibinfo {author} {\bibfnamefont {B.}~\bibnamefont {Keimer}}, \bibinfo {author} {\bibfnamefont {N.}~\bibnamefont {Nagaosa}},\ and\ \bibinfo {author} {\bibfnamefont {Y.}~\bibnamefont {Tokura}},\ }\bibfield  {title} {\enquote {\bibinfo {title} {{Emergent phenomena at oxide interfaces}},}\ }\href {https://doi.org/10.1038/nmat3223} {\bibfield  {journal} {\bibinfo  {journal} {Nature Materials 2012 11:2}\ } (\bibinfo {year} {2012}),\ 10.1038/nmat3223}\BibitemShut {NoStop}%
\bibitem [{\citenamefont {Shi}\ \emph {et~al.}(2023)\citenamefont {Shi}, \citenamefont {Xi}, \citenamefont {Cao}, \citenamefont {Lin}, \citenamefont {Liu}, \citenamefont {Niu}, \citenamefont {Lan}, \citenamefont {Zhou}, \citenamefont {Cao}, \citenamefont {Su}, \citenamefont {Zhao}, \citenamefont {Yang}, \citenamefont {Zhu}, \citenamefont {Yan}, \citenamefont {Tsymbal}, \citenamefont {Tian},\ and\ \citenamefont {Chen}}]{Shi2023-au}%
  \BibitemOpen
  \bibfield  {author} {\bibinfo {author} {\bibfnamefont {S.}~\bibnamefont {Shi}}, \bibinfo {author} {\bibfnamefont {H.}~\bibnamefont {Xi}}, \bibinfo {author} {\bibfnamefont {T.}~\bibnamefont {Cao}}, \bibinfo {author} {\bibfnamefont {W.}~\bibnamefont {Lin}}, \bibinfo {author} {\bibfnamefont {Z.}~\bibnamefont {Liu}}, \bibinfo {author} {\bibfnamefont {J.}~\bibnamefont {Niu}}, \bibinfo {author} {\bibfnamefont {D.}~\bibnamefont {Lan}}, \bibinfo {author} {\bibfnamefont {C.}~\bibnamefont {Zhou}}, \bibinfo {author} {\bibfnamefont {J.}~\bibnamefont {Cao}}, \bibinfo {author} {\bibfnamefont {H.}~\bibnamefont {Su}}, \bibinfo {author} {\bibfnamefont {T.}~\bibnamefont {Zhao}}, \bibinfo {author} {\bibfnamefont {P.}~\bibnamefont {Yang}}, \bibinfo {author} {\bibfnamefont {Y.}~\bibnamefont {Zhu}}, \bibinfo {author} {\bibfnamefont {X.}~\bibnamefont {Yan}}, \bibinfo {author} {\bibfnamefont {E.~Y.}\ \bibnamefont {Tsymbal}}, \bibinfo {author} {\bibfnamefont {H.}~\bibnamefont {Tian}},\ and\ \bibinfo {author} {\bibfnamefont
  {J.}~\bibnamefont {Chen}},\ }\bibfield  {title} {\enquote {\bibinfo {title} {{Interface-engineered ferroelectricity of epitaxial Hf0.5Zr0.5O2 thin films}},}\ }\href {https://doi.org/10.1038/s41467-023-37560-3} {\bibfield  {journal} {\bibinfo  {journal} {Nat. Commun.}\ } (\bibinfo {year} {2023}),\ 10.1038/s41467-023-37560-3}\BibitemShut {NoStop}%
\bibitem [{\citenamefont {Nelson}\ \emph {et~al.}(2011{\natexlab{a}})\citenamefont {Nelson}, \citenamefont {Winchester}, \citenamefont {Zhang}, \citenamefont {Kim}, \citenamefont {Melville}, \citenamefont {Adamo}, \citenamefont {Folkman}, \citenamefont {Baek}, \citenamefont {Eom}, \citenamefont {Schlom}, \citenamefont {Chen},\ and\ \citenamefont {Pan}}]{Nelson2011-zd}%
  \BibitemOpen
  \bibfield  {author} {\bibinfo {author} {\bibfnamefont {C.~T.}\ \bibnamefont {Nelson}}, \bibinfo {author} {\bibfnamefont {B.}~\bibnamefont {Winchester}}, \bibinfo {author} {\bibfnamefont {Y.}~\bibnamefont {Zhang}}, \bibinfo {author} {\bibfnamefont {S.~J.}\ \bibnamefont {Kim}}, \bibinfo {author} {\bibfnamefont {A.}~\bibnamefont {Melville}}, \bibinfo {author} {\bibfnamefont {C.}~\bibnamefont {Adamo}}, \bibinfo {author} {\bibfnamefont {C.~M.}\ \bibnamefont {Folkman}}, \bibinfo {author} {\bibfnamefont {S.~H.}\ \bibnamefont {Baek}}, \bibinfo {author} {\bibfnamefont {C.~B.}\ \bibnamefont {Eom}}, \bibinfo {author} {\bibfnamefont {D.~G.}\ \bibnamefont {Schlom}}, \bibinfo {author} {\bibfnamefont {L.~Q.}\ \bibnamefont {Chen}},\ and\ \bibinfo {author} {\bibfnamefont {X.}~\bibnamefont {Pan}},\ }\bibfield  {title} {\enquote {\bibinfo {title} {{Spontaneous vortex nanodomain arrays at ferroelectric heterointerfaces}},}\ }\href {https://doi.org/10.1021/NL1041808/SUPPL\_FILE/NL1041808\_SI\_001.PDF} {\bibfield  {journal}
  {\bibinfo  {journal} {Nano Lett.}\ } (\bibinfo {year} {2011}{\natexlab{a}}),\ 10.1021/NL1041808/SUPPL\_FILE/NL1041808\_SI\_001.PDF}\BibitemShut {NoStop}%
\bibitem [{\citenamefont {Yadav}\ \emph {et~al.}(2016)\citenamefont {Yadav}, \citenamefont {Nelson}, \citenamefont {Hsu}, \citenamefont {Hong}, \citenamefont {Clarkson}, \citenamefont {Schlepu{\"{e}}tz}, \citenamefont {Damodaran}, \citenamefont {Shafer}, \citenamefont {Arenholz}, \citenamefont {Dedon}, \citenamefont {Chen}, \citenamefont {Vishwanath}, \citenamefont {Minor}, \citenamefont {Chen}, \citenamefont {Scott}, \citenamefont {Martin}, \citenamefont {Ramesh}, \citenamefont {Schlep{\"{u}}tz}, \citenamefont {Damodaran}, \citenamefont {Shafer}, \citenamefont {Arenholz}, \citenamefont {Dedon}, \citenamefont {Chen}, \citenamefont {Vishwanath}, \citenamefont {Minor}, \citenamefont {Chen}, \citenamefont {Scott}, \citenamefont {Martin},\ and\ \citenamefont {Ramesh}}]{Yadav2016-ob}%
  \BibitemOpen
  \bibfield  {author} {\bibinfo {author} {\bibfnamefont {A.~K.}\ \bibnamefont {Yadav}}, \bibinfo {author} {\bibfnamefont {C.~T.}\ \bibnamefont {Nelson}}, \bibinfo {author} {\bibfnamefont {S.~L.}\ \bibnamefont {Hsu}}, \bibinfo {author} {\bibfnamefont {Z.}~\bibnamefont {Hong}}, \bibinfo {author} {\bibfnamefont {J.~D.}\ \bibnamefont {Clarkson}}, \bibinfo {author} {\bibfnamefont {C.~M.}\ \bibnamefont {Schlepu{\"{e}}tz}}, \bibinfo {author} {\bibfnamefont {A.~R.}\ \bibnamefont {Damodaran}}, \bibinfo {author} {\bibfnamefont {P.}~\bibnamefont {Shafer}}, \bibinfo {author} {\bibfnamefont {E.}~\bibnamefont {Arenholz}}, \bibinfo {author} {\bibfnamefont {L.~R.}\ \bibnamefont {Dedon}}, \bibinfo {author} {\bibfnamefont {D.}~\bibnamefont {Chen}}, \bibinfo {author} {\bibfnamefont {A.}~\bibnamefont {Vishwanath}}, \bibinfo {author} {\bibfnamefont {A.~M.}\ \bibnamefont {Minor}}, \bibinfo {author} {\bibfnamefont {L.~Q.}\ \bibnamefont {Chen}}, \bibinfo {author} {\bibfnamefont {J.~F.}\ \bibnamefont {Scott}}, \bibinfo {author}
  {\bibfnamefont {L.~W.}\ \bibnamefont {Martin}}, \bibinfo {author} {\bibfnamefont {R.}~\bibnamefont {Ramesh}}, \bibinfo {author} {\bibfnamefont {C.~M.}\ \bibnamefont {Schlep{\"{u}}tz}}, \bibinfo {author} {\bibfnamefont {A.~R.}\ \bibnamefont {Damodaran}}, \bibinfo {author} {\bibfnamefont {P.}~\bibnamefont {Shafer}}, \bibinfo {author} {\bibfnamefont {E.}~\bibnamefont {Arenholz}}, \bibinfo {author} {\bibfnamefont {L.~R.}\ \bibnamefont {Dedon}}, \bibinfo {author} {\bibfnamefont {D.}~\bibnamefont {Chen}}, \bibinfo {author} {\bibfnamefont {A.}~\bibnamefont {Vishwanath}}, \bibinfo {author} {\bibfnamefont {A.~M.}\ \bibnamefont {Minor}}, \bibinfo {author} {\bibfnamefont {L.~Q.}\ \bibnamefont {Chen}}, \bibinfo {author} {\bibfnamefont {J.~F.}\ \bibnamefont {Scott}}, \bibinfo {author} {\bibfnamefont {L.~W.}\ \bibnamefont {Martin}},\ and\ \bibinfo {author} {\bibfnamefont {R.}~\bibnamefont {Ramesh}},\ }\bibfield  {title} {\enquote {\bibinfo {title} {{Observation of polar vortices in oxide superlattices}},}\ }\href
  {https://doi.org/10.1038/nature16463} {\bibfield  {journal} {\bibinfo  {journal} {Nature}\ } (\bibinfo {year} {2016}),\ 10.1038/nature16463},\ \Eprint {https://arxiv.org/abs/10.1038/nature16463} {10.1038/nature16463} \BibitemShut {NoStop}%
\bibitem [{\citenamefont {Abid}\ \emph {et~al.}(2021)\citenamefont {Abid}, \citenamefont {Sun}, \citenamefont {Hou}, \citenamefont {Tan}, \citenamefont {Zhong}, \citenamefont {Zhu}, \citenamefont {Chen}, \citenamefont {Qu}, \citenamefont {Li}, \citenamefont {Wu}, \citenamefont {Zhang}, \citenamefont {Wang}, \citenamefont {Liu}, \citenamefont {Bai}, \citenamefont {Yu}, \citenamefont {Ouyang}, \citenamefont {Wang}, \citenamefont {Li},\ and\ \citenamefont {Gao}}]{Abid2021-pe}%
  \BibitemOpen
  \bibfield  {author} {\bibinfo {author} {\bibfnamefont {A.~Y.}\ \bibnamefont {Abid}}, \bibinfo {author} {\bibfnamefont {Y.}~\bibnamefont {Sun}}, \bibinfo {author} {\bibfnamefont {X.}~\bibnamefont {Hou}}, \bibinfo {author} {\bibfnamefont {C.}~\bibnamefont {Tan}}, \bibinfo {author} {\bibfnamefont {X.}~\bibnamefont {Zhong}}, \bibinfo {author} {\bibfnamefont {R.}~\bibnamefont {Zhu}}, \bibinfo {author} {\bibfnamefont {H.}~\bibnamefont {Chen}}, \bibinfo {author} {\bibfnamefont {K.}~\bibnamefont {Qu}}, \bibinfo {author} {\bibfnamefont {Y.}~\bibnamefont {Li}}, \bibinfo {author} {\bibfnamefont {M.}~\bibnamefont {Wu}}, \bibinfo {author} {\bibfnamefont {J.}~\bibnamefont {Zhang}}, \bibinfo {author} {\bibfnamefont {J.}~\bibnamefont {Wang}}, \bibinfo {author} {\bibfnamefont {K.}~\bibnamefont {Liu}}, \bibinfo {author} {\bibfnamefont {X.}~\bibnamefont {Bai}}, \bibinfo {author} {\bibfnamefont {D.}~\bibnamefont {Yu}}, \bibinfo {author} {\bibfnamefont {X.}~\bibnamefont {Ouyang}}, \bibinfo {author} {\bibfnamefont
  {J.}~\bibnamefont {Wang}}, \bibinfo {author} {\bibfnamefont {J.}~\bibnamefont {Li}},\ and\ \bibinfo {author} {\bibfnamefont {P.}~\bibnamefont {Gao}},\ }\bibfield  {title} {\enquote {\bibinfo {title} {{Creating polar antivortex in PbTiO3/SrTiO3 superlattice}},}\ }\href {https://doi.org/10.1038/s41467-021-22356-0} {\bibfield  {journal} {\bibinfo  {journal} {Nat. Commun.}\ } (\bibinfo {year} {2021}),\ 10.1038/s41467-021-22356-0}\BibitemShut {NoStop}%
\bibitem [{\citenamefont {Zhu}\ \emph {et~al.}(2025)\citenamefont {Zhu}, \citenamefont {Lanier}, \citenamefont {Genlik}, \citenamefont {Flores}, \citenamefont {Barbosa}, \citenamefont {Randeria}, \citenamefont {Woodward}, \citenamefont {Ghazisaeidi}, \citenamefont {Yang},\ and\ \citenamefont {Hwang}}]{Zhu2025-rm}%
  \BibitemOpen
  \bibfield  {author} {\bibinfo {author} {\bibfnamefont {M.}~\bibnamefont {Zhu}}, \bibinfo {author} {\bibfnamefont {J.}~\bibnamefont {Lanier}}, \bibinfo {author} {\bibfnamefont {S.~P.}\ \bibnamefont {Genlik}}, \bibinfo {author} {\bibfnamefont {J.~G.}\ \bibnamefont {Flores}}, \bibinfo {author} {\bibfnamefont {V.~d. C.~P.}\ \bibnamefont {Barbosa}}, \bibinfo {author} {\bibfnamefont {M.}~\bibnamefont {Randeria}}, \bibinfo {author} {\bibfnamefont {P.~M.}\ \bibnamefont {Woodward}}, \bibinfo {author} {\bibfnamefont {M.}~\bibnamefont {Ghazisaeidi}}, \bibinfo {author} {\bibfnamefont {F.}~\bibnamefont {Yang}},\ and\ \bibinfo {author} {\bibfnamefont {J.}~\bibnamefont {Hwang}},\ }\bibfield  {title} {\enquote {\bibinfo {title} {{Emergent ferromagnetism at {LaFeO}$_{3}$/{SrTiO}$_{3}$ interface arising from a strain-induced spin-state transition}},}\ }\href {https://doi.org/10.1002/admi.202500169} {\bibfield  {journal} {\bibinfo  {journal} {Adv. Mater. Interfaces}\ } (\bibinfo {year} {2025}),\
  10.1002/admi.202500169}\BibitemShut {NoStop}%
\bibitem [{\citenamefont {Xie}\ \emph {et~al.}(2017)\citenamefont {Xie}, \citenamefont {Li}, \citenamefont {Heikes}, \citenamefont {Zhang}, \citenamefont {Hong}, \citenamefont {Gao}, \citenamefont {Nelson}, \citenamefont {Xue}, \citenamefont {Kioupakis}, \citenamefont {Chen}, \citenamefont {Schlom}, \citenamefont {Wang},\ and\ \citenamefont {Pan}}]{Xie2017-zb}%
  \BibitemOpen
  \bibfield  {author} {\bibinfo {author} {\bibfnamefont {L.}~\bibnamefont {Xie}}, \bibinfo {author} {\bibfnamefont {L.}~\bibnamefont {Li}}, \bibinfo {author} {\bibfnamefont {C.~A.}\ \bibnamefont {Heikes}}, \bibinfo {author} {\bibfnamefont {Y.}~\bibnamefont {Zhang}}, \bibinfo {author} {\bibfnamefont {Z.}~\bibnamefont {Hong}}, \bibinfo {author} {\bibfnamefont {P.}~\bibnamefont {Gao}}, \bibinfo {author} {\bibfnamefont {C.~T.}\ \bibnamefont {Nelson}}, \bibinfo {author} {\bibfnamefont {F.}~\bibnamefont {Xue}}, \bibinfo {author} {\bibfnamefont {E.}~\bibnamefont {Kioupakis}}, \bibinfo {author} {\bibfnamefont {L.}~\bibnamefont {Chen}}, \bibinfo {author} {\bibfnamefont {D.~G.}\ \bibnamefont {Schlom}}, \bibinfo {author} {\bibfnamefont {P.}~\bibnamefont {Wang}},\ and\ \bibinfo {author} {\bibfnamefont {X.}~\bibnamefont {Pan}},\ }\bibfield  {title} {\enquote {\bibinfo {title} {{Giant ferroelectric polarization in ultrathin ferroelectrics via boundary-condition engineering}},}\ }\href
  {https://doi.org/10.1002/adma.201701475} {\bibfield  {journal} {\bibinfo  {journal} {Adv. Mater.}\ } (\bibinfo {year} {2017}),\ 10.1002/adma.201701475}\BibitemShut {NoStop}%
\bibitem [{\citenamefont {Junquera}\ and\ \citenamefont {Ghosez}(2003)}]{Junquera2003-ip}%
  \BibitemOpen
  \bibfield  {author} {\bibinfo {author} {\bibfnamefont {J.}~\bibnamefont {Junquera}}\ and\ \bibinfo {author} {\bibfnamefont {P.}~\bibnamefont {Ghosez}},\ }\bibfield  {title} {\enquote {\bibinfo {title} {{Critical thickness for ferroelectricity in perovskite ultrathin films}},}\ }\href {https://doi.org/10.1038/nature01501} {\bibfield  {journal} {\bibinfo  {journal} {Nature}\ } (\bibinfo {year} {2003}),\ 10.1038/nature01501}\BibitemShut {NoStop}%
\bibitem [{\citenamefont {Tenne}\ \emph {et~al.}(2009)\citenamefont {Tenne}, \citenamefont {Turner}, \citenamefont {Schmidt}, \citenamefont {Biegalski}, \citenamefont {Li}, \citenamefont {Chen}, \citenamefont {Soukiassian}, \citenamefont {Trolier-McKinstry}, \citenamefont {Schlom}, \citenamefont {Xi}, \citenamefont {Fong}, \citenamefont {Fuoss}, \citenamefont {Eastman}, \citenamefont {Stephenson}, \citenamefont {Thompson},\ and\ \citenamefont {Streiffer}}]{Tenne2009-ka}%
  \BibitemOpen
  \bibfield  {author} {\bibinfo {author} {\bibfnamefont {D.~A.}\ \bibnamefont {Tenne}}, \bibinfo {author} {\bibfnamefont {P.}~\bibnamefont {Turner}}, \bibinfo {author} {\bibfnamefont {J.~D.}\ \bibnamefont {Schmidt}}, \bibinfo {author} {\bibfnamefont {M.}~\bibnamefont {Biegalski}}, \bibinfo {author} {\bibfnamefont {Y.~L.}\ \bibnamefont {Li}}, \bibinfo {author} {\bibfnamefont {L.~Q.}\ \bibnamefont {Chen}}, \bibinfo {author} {\bibfnamefont {A.}~\bibnamefont {Soukiassian}}, \bibinfo {author} {\bibfnamefont {S.}~\bibnamefont {Trolier-McKinstry}}, \bibinfo {author} {\bibfnamefont {D.~G.}\ \bibnamefont {Schlom}}, \bibinfo {author} {\bibfnamefont {X.~X.}\ \bibnamefont {Xi}}, \bibinfo {author} {\bibfnamefont {D.~D.}\ \bibnamefont {Fong}}, \bibinfo {author} {\bibfnamefont {P.~H.}\ \bibnamefont {Fuoss}}, \bibinfo {author} {\bibfnamefont {J.~A.}\ \bibnamefont {Eastman}}, \bibinfo {author} {\bibfnamefont {G.~B.}\ \bibnamefont {Stephenson}}, \bibinfo {author} {\bibfnamefont {C.}~\bibnamefont {Thompson}},\ and\ \bibinfo
  {author} {\bibfnamefont {S.~K.}\ \bibnamefont {Streiffer}},\ }\bibfield  {title} {\enquote {\bibinfo {title} {{Ferroelectricity in ultrathin BaTiO3 films: probing the size effect by ultraviolet Raman spectroscopy}},}\ }\href {https://doi.org/10.1103/PhysRevLett.103.177601} {\bibfield  {journal} {\bibinfo  {journal} {Phys. Rev. Lett.}\ } (\bibinfo {year} {2009}),\ 10.1103/PhysRevLett.103.177601}\BibitemShut {NoStop}%
\bibitem [{\citenamefont {Lee}\ \emph {et~al.}(2015)\citenamefont {Lee}, \citenamefont {Lu}, \citenamefont {Gu}, \citenamefont {Choi}, \citenamefont {Li}, \citenamefont {Ryu}, \citenamefont {Paudel}, \citenamefont {Song}, \citenamefont {Mikheev}, \citenamefont {Lee}, \citenamefont {Stemmer}, \citenamefont {Tenne}, \citenamefont {Oh}, \citenamefont {Tsymbal}, \citenamefont {Wu}, \citenamefont {Chen}, \citenamefont {Gruverman},\ and\ \citenamefont {Eom}}]{Lee2015-au}%
  \BibitemOpen
  \bibfield  {author} {\bibinfo {author} {\bibfnamefont {D.}~\bibnamefont {Lee}}, \bibinfo {author} {\bibfnamefont {H.}~\bibnamefont {Lu}}, \bibinfo {author} {\bibfnamefont {Y.}~\bibnamefont {Gu}}, \bibinfo {author} {\bibfnamefont {S.-Y.}\ \bibnamefont {Choi}}, \bibinfo {author} {\bibfnamefont {S.-D.}\ \bibnamefont {Li}}, \bibinfo {author} {\bibfnamefont {S.}~\bibnamefont {Ryu}}, \bibinfo {author} {\bibfnamefont {T.~R.}\ \bibnamefont {Paudel}}, \bibinfo {author} {\bibfnamefont {K.}~\bibnamefont {Song}}, \bibinfo {author} {\bibfnamefont {E.}~\bibnamefont {Mikheev}}, \bibinfo {author} {\bibfnamefont {S.}~\bibnamefont {Lee}}, \bibinfo {author} {\bibfnamefont {S.}~\bibnamefont {Stemmer}}, \bibinfo {author} {\bibfnamefont {D.~A.}\ \bibnamefont {Tenne}}, \bibinfo {author} {\bibfnamefont {S.~H.}\ \bibnamefont {Oh}}, \bibinfo {author} {\bibfnamefont {E.~Y.}\ \bibnamefont {Tsymbal}}, \bibinfo {author} {\bibfnamefont {X.}~\bibnamefont {Wu}}, \bibinfo {author} {\bibfnamefont {L.-Q.}\ \bibnamefont {Chen}}, \bibinfo
  {author} {\bibfnamefont {A.}~\bibnamefont {Gruverman}},\ and\ \bibinfo {author} {\bibfnamefont {C.~B.}\ \bibnamefont {Eom}},\ }\bibfield  {title} {\enquote {\bibinfo {title} {{Emergence of room-temperature ferroelectricity at reduced dimensions}},}\ }\href {https://doi.org/10.1126/science.aaa6442} {\bibfield  {journal} {\bibinfo  {journal} {Science}\ } (\bibinfo {year} {2015}),\ 10.1126/science.aaa6442}\BibitemShut {NoStop}%
\bibitem [{\citenamefont {Maksymovych}\ \emph {et~al.}(2012)\citenamefont {Maksymovych}, \citenamefont {Huijben}, \citenamefont {Pan}, \citenamefont {Jesse}, \citenamefont {Balke}, \citenamefont {Chu}, \citenamefont {Chang}, \citenamefont {Borisevich}, \citenamefont {Baddorf}, \citenamefont {Rijnders}, \citenamefont {Blank}, \citenamefont {Ramesh},\ and\ \citenamefont {Kalinin}}]{Maksymovych2012-ul}%
  \BibitemOpen
  \bibfield  {author} {\bibinfo {author} {\bibfnamefont {P.}~\bibnamefont {Maksymovych}}, \bibinfo {author} {\bibfnamefont {M.}~\bibnamefont {Huijben}}, \bibinfo {author} {\bibfnamefont {M.}~\bibnamefont {Pan}}, \bibinfo {author} {\bibfnamefont {S.}~\bibnamefont {Jesse}}, \bibinfo {author} {\bibfnamefont {N.}~\bibnamefont {Balke}}, \bibinfo {author} {\bibfnamefont {Y.-H.}\ \bibnamefont {Chu}}, \bibinfo {author} {\bibfnamefont {H.~J.}\ \bibnamefont {Chang}}, \bibinfo {author} {\bibfnamefont {A.~Y.}\ \bibnamefont {Borisevich}}, \bibinfo {author} {\bibfnamefont {A.~P.}\ \bibnamefont {Baddorf}}, \bibinfo {author} {\bibfnamefont {G.}~\bibnamefont {Rijnders}}, \bibinfo {author} {\bibfnamefont {D.~H.~A.}\ \bibnamefont {Blank}}, \bibinfo {author} {\bibfnamefont {R.}~\bibnamefont {Ramesh}},\ and\ \bibinfo {author} {\bibfnamefont {S.~V.}\ \bibnamefont {Kalinin}},\ }\bibfield  {title} {\enquote {\bibinfo {title} {{Ultrathin limit and dead-layer effects in local polarization switching of {BiFeO3}}},}\ }\href
  {https://doi.org/10.1103/PhysRevB.85.014119} {\bibfield  {journal} {\bibinfo  {journal} {Phys. Rev. B}\ } (\bibinfo {year} {2012}),\ 10.1103/PhysRevB.85.014119}\BibitemShut {NoStop}%
\bibitem [{\citenamefont {Kim}\ \emph {et~al.}(2005)\citenamefont {Kim}, \citenamefont {Kim}, \citenamefont {Kim}, \citenamefont {Chang}, \citenamefont {Noh}, \citenamefont {Kong}, \citenamefont {Char}, \citenamefont {Park}, \citenamefont {Bu}, \citenamefont {Yoon},\ and\ \citenamefont {Chung}}]{Kim2005-sj}%
  \BibitemOpen
  \bibfield  {author} {\bibinfo {author} {\bibfnamefont {Y.~S.}\ \bibnamefont {Kim}}, \bibinfo {author} {\bibfnamefont {D.~H.}\ \bibnamefont {Kim}}, \bibinfo {author} {\bibfnamefont {J.~D.}\ \bibnamefont {Kim}}, \bibinfo {author} {\bibfnamefont {Y.~J.}\ \bibnamefont {Chang}}, \bibinfo {author} {\bibfnamefont {T.~W.}\ \bibnamefont {Noh}}, \bibinfo {author} {\bibfnamefont {J.~H.}\ \bibnamefont {Kong}}, \bibinfo {author} {\bibfnamefont {K.}~\bibnamefont {Char}}, \bibinfo {author} {\bibfnamefont {Y.~D.}\ \bibnamefont {Park}}, \bibinfo {author} {\bibfnamefont {S.~D.}\ \bibnamefont {Bu}}, \bibinfo {author} {\bibfnamefont {J.-G.}\ \bibnamefont {Yoon}},\ and\ \bibinfo {author} {\bibfnamefont {J.-S.}\ \bibnamefont {Chung}},\ }\bibfield  {title} {\enquote {\bibinfo {title} {{Critical thickness of ultrathin ferroelectric BaTiO3 films}},}\ }\href {https://doi.org/10.1063/1.1880443} {\bibfield  {journal} {\bibinfo  {journal} {Appl. Phys. Lett.}\ } (\bibinfo {year} {2005}),\ 10.1063/1.1880443}\BibitemShut {NoStop}%
\bibitem [{\citenamefont {Chang}\ \emph {et~al.}(2011)\citenamefont {Chang}, \citenamefont {Chang}, \citenamefont {Jang}, \citenamefont {Jeong}, \citenamefont {Jung}, \citenamefont {Kim}, \citenamefont {Chung},\ and\ \citenamefont {Noh}}]{Chang2011-xe}%
  \BibitemOpen
  \bibfield  {author} {\bibinfo {author} {\bibfnamefont {S.~H.}\ \bibnamefont {Chang}}, \bibinfo {author} {\bibfnamefont {Y.~J.}\ \bibnamefont {Chang}}, \bibinfo {author} {\bibfnamefont {S.~Y.}\ \bibnamefont {Jang}}, \bibinfo {author} {\bibfnamefont {D.~W.}\ \bibnamefont {Jeong}}, \bibinfo {author} {\bibfnamefont {C.~U.}\ \bibnamefont {Jung}}, \bibinfo {author} {\bibfnamefont {Y.-J.}\ \bibnamefont {Kim}}, \bibinfo {author} {\bibfnamefont {J.-S.}\ \bibnamefont {Chung}},\ and\ \bibinfo {author} {\bibfnamefont {T.~W.}\ \bibnamefont {Noh}},\ }\bibfield  {title} {\enquote {\bibinfo {title} {{Thickness-dependent structural phase transition of strained SrRuO3ultrathin films: The role of octahedral tilt}},}\ }\href {https://doi.org/10.1103/physrevb.84.104101} {\bibfield  {journal} {\bibinfo  {journal} {Phys. Rev. B Condens. Matter Mater. Phys.}\ } (\bibinfo {year} {2011}),\ 10.1103/physrevb.84.104101}\BibitemShut {NoStop}%
\bibitem [{\citenamefont {Zhu*}\ \emph {et~al.}(2025)\citenamefont {Zhu*}, \citenamefont {Xu*}, \citenamefont {Yun}, \citenamefont {Wu}, \citenamefont {Shafir}, \citenamefont {Gilgenbach}, \citenamefont {Martin}, \citenamefont {Grinberg}, \citenamefont {Spanier},\ and\ \citenamefont {LeBeau}}]{Zhu*2025-ub}%
  \BibitemOpen
  \bibfield  {author} {\bibinfo {author} {\bibfnamefont {M.}~\bibnamefont {Zhu*}}, \bibinfo {author} {\bibfnamefont {M.}~\bibnamefont {Xu*}}, \bibinfo {author} {\bibfnamefont {Y.}~\bibnamefont {Yun}}, \bibinfo {author} {\bibfnamefont {L.}~\bibnamefont {Wu}}, \bibinfo {author} {\bibfnamefont {O.}~\bibnamefont {Shafir}}, \bibinfo {author} {\bibfnamefont {C.}~\bibnamefont {Gilgenbach}}, \bibinfo {author} {\bibfnamefont {L.~W.}\ \bibnamefont {Martin}}, \bibinfo {author} {\bibfnamefont {I.}~\bibnamefont {Grinberg}}, \bibinfo {author} {\bibfnamefont {J.~E.}\ \bibnamefont {Spanier}},\ and\ \bibinfo {author} {\bibfnamefont {J.~M.}\ \bibnamefont {LeBeau}},\ }\bibfield  {title} {\enquote {\bibinfo {title} {{Insights into chemical and structural order at planar defects in Pb2MgWO6 using multislice electron ptychography}},}\ }\href {https://doi.org/10.1021/acsnano.4c14833} {\bibfield  {journal} {\bibinfo  {journal} {ACS Nano}\ } (\bibinfo {year} {2025}),\ 10.1021/acsnano.4c14833}\BibitemShut {NoStop}%
\bibitem [{\citenamefont {Li}, \citenamefont {Xie},\ and\ \citenamefont {Pan}(2019)}]{Li2019-en}%
  \BibitemOpen
  \bibfield  {author} {\bibinfo {author} {\bibfnamefont {L.}~\bibnamefont {Li}}, \bibinfo {author} {\bibfnamefont {L.}~\bibnamefont {Xie}},\ and\ \bibinfo {author} {\bibfnamefont {X.}~\bibnamefont {Pan}},\ }\bibfield  {title} {\enquote {\bibinfo {title} {{Real-time studies of ferroelectric domain switching: a review}},}\ }\href {https://doi.org/10.1088/1361-6633/ab28de} {\bibfield  {journal} {\bibinfo  {journal} {Rep. Prog. Phys.}\ } (\bibinfo {year} {2019}),\ 10.1088/1361-6633/ab28de}\BibitemShut {NoStop}%
\bibitem [{\citenamefont {Nelson}\ \emph {et~al.}(2011{\natexlab{b}})\citenamefont {Nelson}, \citenamefont {Gao}, \citenamefont {Jokisaari}, \citenamefont {Heikes}, \citenamefont {Adamo}, \citenamefont {Melville}, \citenamefont {Baek}, \citenamefont {Folkman}, \citenamefont {Winchester}, \citenamefont {Gu}, \citenamefont {Liu}, \citenamefont {Zhang}, \citenamefont {Wang}, \citenamefont {Li}, \citenamefont {Chen}, \citenamefont {Eom}, \citenamefont {Schlom},\ and\ \citenamefont {Pan}}]{Nelson2011-gk}%
  \BibitemOpen
  \bibfield  {author} {\bibinfo {author} {\bibfnamefont {C.~T.}\ \bibnamefont {Nelson}}, \bibinfo {author} {\bibfnamefont {P.}~\bibnamefont {Gao}}, \bibinfo {author} {\bibfnamefont {J.~R.}\ \bibnamefont {Jokisaari}}, \bibinfo {author} {\bibfnamefont {C.}~\bibnamefont {Heikes}}, \bibinfo {author} {\bibfnamefont {C.}~\bibnamefont {Adamo}}, \bibinfo {author} {\bibfnamefont {A.}~\bibnamefont {Melville}}, \bibinfo {author} {\bibfnamefont {S.-H.}\ \bibnamefont {Baek}}, \bibinfo {author} {\bibfnamefont {C.~M.}\ \bibnamefont {Folkman}}, \bibinfo {author} {\bibfnamefont {B.}~\bibnamefont {Winchester}}, \bibinfo {author} {\bibfnamefont {Y.}~\bibnamefont {Gu}}, \bibinfo {author} {\bibfnamefont {Y.}~\bibnamefont {Liu}}, \bibinfo {author} {\bibfnamefont {K.}~\bibnamefont {Zhang}}, \bibinfo {author} {\bibfnamefont {E.}~\bibnamefont {Wang}}, \bibinfo {author} {\bibfnamefont {J.}~\bibnamefont {Li}}, \bibinfo {author} {\bibfnamefont {L.-Q.}\ \bibnamefont {Chen}}, \bibinfo {author} {\bibfnamefont {C.-B.}\ \bibnamefont {Eom}},
  \bibinfo {author} {\bibfnamefont {D.~G.}\ \bibnamefont {Schlom}},\ and\ \bibinfo {author} {\bibfnamefont {X.}~\bibnamefont {Pan}},\ }\bibfield  {title} {\enquote {\bibinfo {title} {{Domain dynamics during ferroelectric switching}},}\ }\href {https://doi.org/10.1126/science.1206980} {\bibfield  {journal} {\bibinfo  {journal} {Science}\ } (\bibinfo {year} {2011}{\natexlab{b}}),\ 10.1126/science.1206980}\BibitemShut {NoStop}%
\bibitem [{\citenamefont {Pan*}\ \emph {et~al.}(2024)\citenamefont {Pan*}, \citenamefont {Zhu*}, \citenamefont {Banyas}, \citenamefont {Alaerts}, \citenamefont {Acharya}, \citenamefont {Zhang}, \citenamefont {Kim}, \citenamefont {Chen}, \citenamefont {Huang}, \citenamefont {Xu}, \citenamefont {Harris}, \citenamefont {Tian}, \citenamefont {Ricci}, \citenamefont {Hanrahan}, \citenamefont {Spanier}, \citenamefont {Hautier}, \citenamefont {LeBeau}, \citenamefont {Neaton},\ and\ \citenamefont {Martin}}]{Pan*2024-xy}%
  \BibitemOpen
  \bibfield  {author} {\bibinfo {author} {\bibfnamefont {H.}~\bibnamefont {Pan*}}, \bibinfo {author} {\bibfnamefont {M.}~\bibnamefont {Zhu*}}, \bibinfo {author} {\bibfnamefont {E.}~\bibnamefont {Banyas}}, \bibinfo {author} {\bibfnamefont {L.}~\bibnamefont {Alaerts}}, \bibinfo {author} {\bibfnamefont {M.}~\bibnamefont {Acharya}}, \bibinfo {author} {\bibfnamefont {H.}~\bibnamefont {Zhang}}, \bibinfo {author} {\bibfnamefont {J.}~\bibnamefont {Kim}}, \bibinfo {author} {\bibfnamefont {X.}~\bibnamefont {Chen}}, \bibinfo {author} {\bibfnamefont {X.}~\bibnamefont {Huang}}, \bibinfo {author} {\bibfnamefont {M.}~\bibnamefont {Xu}}, \bibinfo {author} {\bibfnamefont {I.}~\bibnamefont {Harris}}, \bibinfo {author} {\bibfnamefont {Z.}~\bibnamefont {Tian}}, \bibinfo {author} {\bibfnamefont {F.}~\bibnamefont {Ricci}}, \bibinfo {author} {\bibfnamefont {B.}~\bibnamefont {Hanrahan}}, \bibinfo {author} {\bibfnamefont {J.~E.}\ \bibnamefont {Spanier}}, \bibinfo {author} {\bibfnamefont {G.}~\bibnamefont {Hautier}}, \bibinfo {author}
  {\bibfnamefont {J.~M.}\ \bibnamefont {LeBeau}}, \bibinfo {author} {\bibfnamefont {J.~B.}\ \bibnamefont {Neaton}},\ and\ \bibinfo {author} {\bibfnamefont {L.~W.}\ \bibnamefont {Martin}},\ }\bibfield  {title} {\enquote {\bibinfo {title} {{Clamping enables enhanced electromechanical responses in antiferroelectric thin films}},}\ }\href {https://doi.org/10.1038/s41563-024-01907-y} {\bibfield  {journal} {\bibinfo  {journal} {Nat. Mater.}\ } (\bibinfo {year} {2024}),\ 10.1038/s41563-024-01907-y}\BibitemShut {NoStop}%
\bibitem [{\citenamefont {Nukala}\ \emph {et~al.}(2021)\citenamefont {Nukala}, \citenamefont {Ahmadi}, \citenamefont {Wei}, \citenamefont {de~Graaf}, \citenamefont {Stylianidis}, \citenamefont {Chakrabortty}, \citenamefont {Matzen}, \citenamefont {Zandbergen}, \citenamefont {Bj{\"{o}}rling}, \citenamefont {Mannix}, \citenamefont {Carbone}, \citenamefont {Kooi},\ and\ \citenamefont {Noheda}}]{Nukala2021-tp}%
  \BibitemOpen
  \bibfield  {author} {\bibinfo {author} {\bibfnamefont {P.}~\bibnamefont {Nukala}}, \bibinfo {author} {\bibfnamefont {M.}~\bibnamefont {Ahmadi}}, \bibinfo {author} {\bibfnamefont {Y.}~\bibnamefont {Wei}}, \bibinfo {author} {\bibfnamefont {S.}~\bibnamefont {de~Graaf}}, \bibinfo {author} {\bibfnamefont {E.}~\bibnamefont {Stylianidis}}, \bibinfo {author} {\bibfnamefont {T.}~\bibnamefont {Chakrabortty}}, \bibinfo {author} {\bibfnamefont {S.}~\bibnamefont {Matzen}}, \bibinfo {author} {\bibfnamefont {H.~W.}\ \bibnamefont {Zandbergen}}, \bibinfo {author} {\bibfnamefont {A.}~\bibnamefont {Bj{\"{o}}rling}}, \bibinfo {author} {\bibfnamefont {D.}~\bibnamefont {Mannix}}, \bibinfo {author} {\bibfnamefont {D.}~\bibnamefont {Carbone}}, \bibinfo {author} {\bibfnamefont {B.}~\bibnamefont {Kooi}},\ and\ \bibinfo {author} {\bibfnamefont {B.}~\bibnamefont {Noheda}},\ }\bibfield  {title} {\enquote {\bibinfo {title} {{{Reversible oxygen migration and phase transitions in hafnia-based ferroelectric devices}}},}\ }\href
  {https://doi.org/10.1126/science.abf3789} {\bibfield  {journal} {\bibinfo  {journal} {Science}\ } (\bibinfo {year} {2021}),\ 10.1126/science.abf3789}\BibitemShut {NoStop}%
\bibitem [{\citenamefont {Zhang}\ \emph {et~al.}(2017)\citenamefont {Zhang}, \citenamefont {He}, \citenamefont {Shi}, \citenamefont {Lu}, \citenamefont {Li}, \citenamefont {Yu}, \citenamefont {Zhang}, \citenamefont {Chen}, \citenamefont {Morris}, \citenamefont {Xu}, \citenamefont {Yu}, \citenamefont {Gu}, \citenamefont {Jin},\ and\ \citenamefont {Nan}}]{Zhang2017-mo}%
  \BibitemOpen
  \bibfield  {author} {\bibinfo {author} {\bibfnamefont {Q.}~\bibnamefont {Zhang}}, \bibinfo {author} {\bibfnamefont {X.}~\bibnamefont {He}}, \bibinfo {author} {\bibfnamefont {J.}~\bibnamefont {Shi}}, \bibinfo {author} {\bibfnamefont {N.}~\bibnamefont {Lu}}, \bibinfo {author} {\bibfnamefont {H.}~\bibnamefont {Li}}, \bibinfo {author} {\bibfnamefont {Q.}~\bibnamefont {Yu}}, \bibinfo {author} {\bibfnamefont {Z.}~\bibnamefont {Zhang}}, \bibinfo {author} {\bibfnamefont {L.-Q.}\ \bibnamefont {Chen}}, \bibinfo {author} {\bibfnamefont {B.}~\bibnamefont {Morris}}, \bibinfo {author} {\bibfnamefont {Q.}~\bibnamefont {Xu}}, \bibinfo {author} {\bibfnamefont {P.}~\bibnamefont {Yu}}, \bibinfo {author} {\bibfnamefont {L.}~\bibnamefont {Gu}}, \bibinfo {author} {\bibfnamefont {K.}~\bibnamefont {Jin}},\ and\ \bibinfo {author} {\bibfnamefont {C.-W.}\ \bibnamefont {Nan}},\ }\bibfield  {title} {\enquote {\bibinfo {title} {{Atomic-resolution imaging of electrically induced oxygen vacancy migration and phase transformation in
  SrCoO2.5-$\sigma$}},}\ }\href {https://doi.org/10.1038/s41467-017-00121-6} {\bibfield  {journal} {\bibinfo  {journal} {Nat. Commun.}\ } (\bibinfo {year} {2017}),\ 10.1038/s41467-017-00121-6}\BibitemShut {NoStop}%
\bibitem [{\citenamefont {Gong}\ \emph {et~al.}(2018)\citenamefont {Gong}, \citenamefont {Chen}, \citenamefont {Zhang}, \citenamefont {Meng}, \citenamefont {Shi}, \citenamefont {Liu}, \citenamefont {Liu}, \citenamefont {Zhang}, \citenamefont {Wang}, \citenamefont {Wang}, \citenamefont {Yu}, \citenamefont {Zhang}, \citenamefont {Xu}, \citenamefont {Xiao}, \citenamefont {Hu}, \citenamefont {Gu}, \citenamefont {Li}, \citenamefont {Huang},\ and\ \citenamefont {Chen}}]{Gong2018-ei}%
  \BibitemOpen
  \bibfield  {author} {\bibinfo {author} {\bibfnamefont {Y.}~\bibnamefont {Gong}}, \bibinfo {author} {\bibfnamefont {Y.}~\bibnamefont {Chen}}, \bibinfo {author} {\bibfnamefont {Q.}~\bibnamefont {Zhang}}, \bibinfo {author} {\bibfnamefont {F.}~\bibnamefont {Meng}}, \bibinfo {author} {\bibfnamefont {J.-A.}\ \bibnamefont {Shi}}, \bibinfo {author} {\bibfnamefont {X.}~\bibnamefont {Liu}}, \bibinfo {author} {\bibfnamefont {X.}~\bibnamefont {Liu}}, \bibinfo {author} {\bibfnamefont {J.}~\bibnamefont {Zhang}}, \bibinfo {author} {\bibfnamefont {H.}~\bibnamefont {Wang}}, \bibinfo {author} {\bibfnamefont {J.}~\bibnamefont {Wang}}, \bibinfo {author} {\bibfnamefont {Q.}~\bibnamefont {Yu}}, \bibinfo {author} {\bibfnamefont {Z.}~\bibnamefont {Zhang}}, \bibinfo {author} {\bibfnamefont {Q.}~\bibnamefont {Xu}}, \bibinfo {author} {\bibfnamefont {R.}~\bibnamefont {Xiao}}, \bibinfo {author} {\bibfnamefont {Y.-S.}\ \bibnamefont {Hu}}, \bibinfo {author} {\bibfnamefont {L.}~\bibnamefont {Gu}}, \bibinfo {author} {\bibfnamefont
  {H.}~\bibnamefont {Li}}, \bibinfo {author} {\bibfnamefont {X.}~\bibnamefont {Huang}},\ and\ \bibinfo {author} {\bibfnamefont {L.}~\bibnamefont {Chen}},\ }\bibfield  {title} {\enquote {\bibinfo {title} {{Three-dimensional atomic-scale observation of structural evolution of cathode material in a working all-solid-state battery}},}\ }\href {https://doi.org/10.1038/s41467-018-05833-x} {\bibfield  {journal} {\bibinfo  {journal} {Nat. Commun.}\ } (\bibinfo {year} {2018}),\ 10.1038/s41467-018-05833-x}\BibitemShut {NoStop}%
\bibitem [{\citenamefont {Recalde-Benitez}\ \emph {et~al.}(2023)\citenamefont {Recalde-Benitez}, \citenamefont {Jiang}, \citenamefont {Winkler}, \citenamefont {Ruan}, \citenamefont {Zintler}, \citenamefont {Adabifiroozjaei}, \citenamefont {Arzumanov}, \citenamefont {Hubbard}, \citenamefont {van Omme}, \citenamefont {Pivak}, \citenamefont {Perez-Garza}, \citenamefont {Regan}, \citenamefont {Alff}, \citenamefont {Komissinskiy},\ and\ \citenamefont {Molina-Luna}}]{Recalde-Benitez2023-vd}%
  \BibitemOpen
  \bibfield  {author} {\bibinfo {author} {\bibfnamefont {O.}~\bibnamefont {Recalde-Benitez}}, \bibinfo {author} {\bibfnamefont {T.}~\bibnamefont {Jiang}}, \bibinfo {author} {\bibfnamefont {R.}~\bibnamefont {Winkler}}, \bibinfo {author} {\bibfnamefont {Y.}~\bibnamefont {Ruan}}, \bibinfo {author} {\bibfnamefont {A.}~\bibnamefont {Zintler}}, \bibinfo {author} {\bibfnamefont {E.}~\bibnamefont {Adabifiroozjaei}}, \bibinfo {author} {\bibfnamefont {A.}~\bibnamefont {Arzumanov}}, \bibinfo {author} {\bibfnamefont {W.~A.}\ \bibnamefont {Hubbard}}, \bibinfo {author} {\bibfnamefont {T.}~\bibnamefont {van Omme}}, \bibinfo {author} {\bibfnamefont {Y.}~\bibnamefont {Pivak}}, \bibinfo {author} {\bibfnamefont {H.~H.}\ \bibnamefont {Perez-Garza}}, \bibinfo {author} {\bibfnamefont {B.~C.}\ \bibnamefont {Regan}}, \bibinfo {author} {\bibfnamefont {L.}~\bibnamefont {Alff}}, \bibinfo {author} {\bibfnamefont {P.}~\bibnamefont {Komissinskiy}},\ and\ \bibinfo {author} {\bibfnamefont {L.}~\bibnamefont {Molina-Luna}},\ }\bibfield
  {title} {\enquote {\bibinfo {title} {{Operando two-terminal devices inside a transmission electron microscope}},}\ }\href {https://doi.org/10.1038/s44172-023-00133-9} {\bibfield  {journal} {\bibinfo  {journal} {Commun Eng}\ } (\bibinfo {year} {2023}),\ 10.1038/s44172-023-00133-9}\BibitemShut {NoStop}%
\bibitem [{\citenamefont {Recalde-Benitez}\ \emph {et~al.}(2024)\citenamefont {Recalde-Benitez}, \citenamefont {Pivak}, \citenamefont {Jiang}, \citenamefont {Winkler}, \citenamefont {Zintler}, \citenamefont {Adabifiroozjaei}, \citenamefont {Komissinskiy}, \citenamefont {Alff}, \citenamefont {Hubbard}, \citenamefont {Perez-Garza},\ and\ \citenamefont {Molina-Luna}}]{Recalde-Benitez2024-tq}%
  \BibitemOpen
  \bibfield  {author} {\bibinfo {author} {\bibfnamefont {O.}~\bibnamefont {Recalde-Benitez}}, \bibinfo {author} {\bibfnamefont {Y.}~\bibnamefont {Pivak}}, \bibinfo {author} {\bibfnamefont {T.}~\bibnamefont {Jiang}}, \bibinfo {author} {\bibfnamefont {R.}~\bibnamefont {Winkler}}, \bibinfo {author} {\bibfnamefont {A.}~\bibnamefont {Zintler}}, \bibinfo {author} {\bibfnamefont {E.}~\bibnamefont {Adabifiroozjaei}}, \bibinfo {author} {\bibfnamefont {P.}~\bibnamefont {Komissinskiy}}, \bibinfo {author} {\bibfnamefont {L.}~\bibnamefont {Alff}}, \bibinfo {author} {\bibfnamefont {W.~A.}\ \bibnamefont {Hubbard}}, \bibinfo {author} {\bibfnamefont {H.~H.}\ \bibnamefont {Perez-Garza}},\ and\ \bibinfo {author} {\bibfnamefont {L.}~\bibnamefont {Molina-Luna}},\ }\bibfield  {title} {\enquote {\bibinfo {title} {{Weld-free mounting of lamellae for electrical biasing operando {TEM}}},}\ }\href {https://doi.org/10.1016/j.ultramic.2024.113939} {\bibfield  {journal} {\bibinfo  {journal} {Ultramicroscopy}\ } (\bibinfo {year} {2024}),\
  10.1016/j.ultramic.2024.113939}\BibitemShut {NoStop}%
\bibitem [{\citenamefont {Han}\ \emph {et~al.}(2014)\citenamefont {Han}, \citenamefont {Marshall}, \citenamefont {Wu}, \citenamefont {Schofield}, \citenamefont {Aoki}, \citenamefont {Twesten}, \citenamefont {Hoffman}, \citenamefont {Walker}, \citenamefont {Ahn},\ and\ \citenamefont {Zhu}}]{Han2014-rz}%
  \BibitemOpen
  \bibfield  {author} {\bibinfo {author} {\bibfnamefont {M.-G.}\ \bibnamefont {Han}}, \bibinfo {author} {\bibfnamefont {M.~S.~J.}\ \bibnamefont {Marshall}}, \bibinfo {author} {\bibfnamefont {L.}~\bibnamefont {Wu}}, \bibinfo {author} {\bibfnamefont {M.~A.}\ \bibnamefont {Schofield}}, \bibinfo {author} {\bibfnamefont {T.}~\bibnamefont {Aoki}}, \bibinfo {author} {\bibfnamefont {R.}~\bibnamefont {Twesten}}, \bibinfo {author} {\bibfnamefont {J.}~\bibnamefont {Hoffman}}, \bibinfo {author} {\bibfnamefont {F.~J.}\ \bibnamefont {Walker}}, \bibinfo {author} {\bibfnamefont {C.~H.}\ \bibnamefont {Ahn}},\ and\ \bibinfo {author} {\bibfnamefont {Y.}~\bibnamefont {Zhu}},\ }\bibfield  {title} {\enquote {\bibinfo {title} {{Interface-induced nonswitchable domains in ferroelectric thin films}},}\ }\href {https://doi.org/10.1038/ncomms5693} {\bibfield  {journal} {\bibinfo  {journal} {Nat. Commun.}\ } (\bibinfo {year} {2014}),\ 10.1038/ncomms5693}\BibitemShut {NoStop}%
\bibitem [{\citenamefont {Morozovska}\ \emph {et~al.}(2018)\citenamefont {Morozovska}, \citenamefont {Eliseev}, \citenamefont {Vorotiahin}, \citenamefont {Silibin}, \citenamefont {Kalinin},\ and\ \citenamefont {Morozovsky}}]{Morozovska2018-ed}%
  \BibitemOpen
  \bibfield  {author} {\bibinfo {author} {\bibfnamefont {A.~N.}\ \bibnamefont {Morozovska}}, \bibinfo {author} {\bibfnamefont {E.~A.}\ \bibnamefont {Eliseev}}, \bibinfo {author} {\bibfnamefont {I.~S.}\ \bibnamefont {Vorotiahin}}, \bibinfo {author} {\bibfnamefont {M.~V.}\ \bibnamefont {Silibin}}, \bibinfo {author} {\bibfnamefont {S.~V.}\ \bibnamefont {Kalinin}},\ and\ \bibinfo {author} {\bibfnamefont {N.~V.}\ \bibnamefont {Morozovsky}},\ }\bibfield  {title} {\enquote {\bibinfo {title} {{Control of polarization reversal temperature behavior by surface screening in thin ferroelectric films}},}\ }\href {https://doi.org/10.1016/j.actamat.2018.08.041} {\bibfield  {journal} {\bibinfo  {journal} {Acta Mater.}\ } (\bibinfo {year} {2018}),\ 10.1016/j.actamat.2018.08.041}\BibitemShut {NoStop}%
\bibitem [{\citenamefont {Tate}\ \emph {et~al.}(2016)\citenamefont {Tate}, \citenamefont {Purohit}, \citenamefont {Chamberlain}, \citenamefont {Nguyen}, \citenamefont {Hovden}, \citenamefont {Chang}, \citenamefont {Deb}, \citenamefont {Turgut}, \citenamefont {Heron}, \citenamefont {Schlom}, \citenamefont {Ralph}, \citenamefont {Fuchs}, \citenamefont {Shanks}, \citenamefont {Philipp}, \citenamefont {Muller},\ and\ \citenamefont {Gruner}}]{Tate2016-kc}%
  \BibitemOpen
  \bibfield  {author} {\bibinfo {author} {\bibfnamefont {M.~W.}\ \bibnamefont {Tate}}, \bibinfo {author} {\bibfnamefont {P.}~\bibnamefont {Purohit}}, \bibinfo {author} {\bibfnamefont {D.}~\bibnamefont {Chamberlain}}, \bibinfo {author} {\bibfnamefont {K.~X.}\ \bibnamefont {Nguyen}}, \bibinfo {author} {\bibfnamefont {R.}~\bibnamefont {Hovden}}, \bibinfo {author} {\bibfnamefont {C.~S.}\ \bibnamefont {Chang}}, \bibinfo {author} {\bibfnamefont {P.}~\bibnamefont {Deb}}, \bibinfo {author} {\bibfnamefont {E.}~\bibnamefont {Turgut}}, \bibinfo {author} {\bibfnamefont {J.~T.}\ \bibnamefont {Heron}}, \bibinfo {author} {\bibfnamefont {D.~G.}\ \bibnamefont {Schlom}}, \bibinfo {author} {\bibfnamefont {D.~C.}\ \bibnamefont {Ralph}}, \bibinfo {author} {\bibfnamefont {G.~D.}\ \bibnamefont {Fuchs}}, \bibinfo {author} {\bibfnamefont {K.~S.}\ \bibnamefont {Shanks}}, \bibinfo {author} {\bibfnamefont {H.~T.}\ \bibnamefont {Philipp}}, \bibinfo {author} {\bibfnamefont {D.~A.}\ \bibnamefont {Muller}},\ and\ \bibinfo {author}
  {\bibfnamefont {S.~M.}\ \bibnamefont {Gruner}},\ }\bibfield  {title} {\enquote {\bibinfo {title} {{High Dynamic Range Pixel Array Detector for Scanning Transmission Electron Microscopy}},}\ }\href {https://doi.org/10.1017/S1431927615015664} {\bibfield  {journal} {\bibinfo  {journal} {Microsc. Microanal.}\ } (\bibinfo {year} {2016}),\ 10.1017/S1431927615015664},\ \Eprint {https://arxiv.org/abs/1511.03539} {1511.03539} \BibitemShut {NoStop}%
\bibitem [{\citenamefont {Padgett}\ \emph {et~al.}(2020)\citenamefont {Padgett}, \citenamefont {Holtz}, \citenamefont {Cueva}, \citenamefont {Shao}, \citenamefont {Langenberg}, \citenamefont {Schlom},\ and\ \citenamefont {Muller}}]{Padgett2020-au}%
  \BibitemOpen
  \bibfield  {author} {\bibinfo {author} {\bibfnamefont {E.}~\bibnamefont {Padgett}}, \bibinfo {author} {\bibfnamefont {M.~E.}\ \bibnamefont {Holtz}}, \bibinfo {author} {\bibfnamefont {P.}~\bibnamefont {Cueva}}, \bibinfo {author} {\bibfnamefont {Y.-T.~T.}\ \bibnamefont {Shao}}, \bibinfo {author} {\bibfnamefont {E.}~\bibnamefont {Langenberg}}, \bibinfo {author} {\bibfnamefont {D.~G.}\ \bibnamefont {Schlom}},\ and\ \bibinfo {author} {\bibfnamefont {D.~A.}\ \bibnamefont {Muller}},\ }\bibfield  {title} {\enquote {\bibinfo {title} {{The exit-wave power-cepstrum transform for scanning nanobeam electron diffraction: robust strain mapping at subnanometer resolution and subpicometer precision}},}\ }\href {https://doi.org/10.1016/j.ultramic.2020.112994} {\bibfield  {journal} {\bibinfo  {journal} {Ultramicroscopy}\ } (\bibinfo {year} {2020}),\ 10.1016/j.ultramic.2020.112994}\BibitemShut {NoStop}%
\bibitem [{\citenamefont {Kukhar}\ \emph {et~al.}(2006)\citenamefont {Kukhar}, \citenamefont {Pertsev}, \citenamefont {Kohlstedt},\ and\ \citenamefont {Waser}}]{Kukhar2006-yd}%
  \BibitemOpen
  \bibfield  {author} {\bibinfo {author} {\bibfnamefont {V.~G.}\ \bibnamefont {Kukhar}}, \bibinfo {author} {\bibfnamefont {N.~A.}\ \bibnamefont {Pertsev}}, \bibinfo {author} {\bibfnamefont {H.}~\bibnamefont {Kohlstedt}},\ and\ \bibinfo {author} {\bibfnamefont {R.}~\bibnamefont {Waser}},\ }\bibfield  {title} {\enquote {\bibinfo {title} {{Polarization states of polydomain epitaxialPb(Zr1-xTix)O3thin films and their dielectric properties}},}\ }\href {https://doi.org/10.1103/physrevb.73.214103} {\bibfield  {journal} {\bibinfo  {journal} {Phys. Rev. B Condens. Matter Mater. Phys.}\ } (\bibinfo {year} {2006}),\ 10.1103/physrevb.73.214103}\BibitemShut {NoStop}%
\bibitem [{\citenamefont {Damodaran}\ \emph {et~al.}(2017)\citenamefont {Damodaran}, \citenamefont {Pandya}, \citenamefont {Agar}, \citenamefont {Cao}, \citenamefont {Vasudevan}, \citenamefont {Xu}, \citenamefont {Saremi}, \citenamefont {Li}, \citenamefont {Kim}, \citenamefont {McCarter}, \citenamefont {Dedon}, \citenamefont {Angsten}, \citenamefont {Balke}, \citenamefont {Jesse}, \citenamefont {Asta}, \citenamefont {Kalinin},\ and\ \citenamefont {Martin}}]{Damodaran2017-kj}%
  \BibitemOpen
  \bibfield  {author} {\bibinfo {author} {\bibfnamefont {A.~R.}\ \bibnamefont {Damodaran}}, \bibinfo {author} {\bibfnamefont {S.}~\bibnamefont {Pandya}}, \bibinfo {author} {\bibfnamefont {J.~C.}\ \bibnamefont {Agar}}, \bibinfo {author} {\bibfnamefont {Y.}~\bibnamefont {Cao}}, \bibinfo {author} {\bibfnamefont {R.~K.}\ \bibnamefont {Vasudevan}}, \bibinfo {author} {\bibfnamefont {R.}~\bibnamefont {Xu}}, \bibinfo {author} {\bibfnamefont {S.}~\bibnamefont {Saremi}}, \bibinfo {author} {\bibfnamefont {Q.}~\bibnamefont {Li}}, \bibinfo {author} {\bibfnamefont {J.}~\bibnamefont {Kim}}, \bibinfo {author} {\bibfnamefont {M.~R.}\ \bibnamefont {McCarter}}, \bibinfo {author} {\bibfnamefont {L.~R.}\ \bibnamefont {Dedon}}, \bibinfo {author} {\bibfnamefont {T.}~\bibnamefont {Angsten}}, \bibinfo {author} {\bibfnamefont {N.}~\bibnamefont {Balke}}, \bibinfo {author} {\bibfnamefont {S.}~\bibnamefont {Jesse}}, \bibinfo {author} {\bibfnamefont {M.}~\bibnamefont {Asta}}, \bibinfo {author} {\bibfnamefont {S.~V.}\ \bibnamefont
  {Kalinin}},\ and\ \bibinfo {author} {\bibfnamefont {L.~W.}\ \bibnamefont {Martin}},\ }\bibfield  {title} {\enquote {\bibinfo {title} {{Three-state ferroelastic switching and large electromechanical responses in {PbTiO}$_{3}$ thin films}},}\ }\href {https://doi.org/10.1002/adma.201702069} {\bibfield  {journal} {\bibinfo  {journal} {Adv. Mater.}\ } (\bibinfo {year} {2017}),\ 10.1002/adma.201702069}\BibitemShut {NoStop}%
\bibitem [{\citenamefont {Lu}\ \emph {et~al.}(2019)\citenamefont {Lu}, \citenamefont {Chen}, \citenamefont {Cao}, \citenamefont {Tang}, \citenamefont {Xu}, \citenamefont {Saremi}, \citenamefont {Zhang}, \citenamefont {You}, \citenamefont {Dong}, \citenamefont {Das}, \citenamefont {Zhang}, \citenamefont {Zheng}, \citenamefont {Wu}, \citenamefont {Lv}, \citenamefont {Xie}, \citenamefont {Liu}, \citenamefont {Li}, \citenamefont {Chen}, \citenamefont {Chen}, \citenamefont {Cao},\ and\ \citenamefont {Martin}}]{Lu2019-cb}%
  \BibitemOpen
  \bibfield  {author} {\bibinfo {author} {\bibfnamefont {X.}~\bibnamefont {Lu}}, \bibinfo {author} {\bibfnamefont {Z.}~\bibnamefont {Chen}}, \bibinfo {author} {\bibfnamefont {Y.}~\bibnamefont {Cao}}, \bibinfo {author} {\bibfnamefont {Y.}~\bibnamefont {Tang}}, \bibinfo {author} {\bibfnamefont {R.}~\bibnamefont {Xu}}, \bibinfo {author} {\bibfnamefont {S.}~\bibnamefont {Saremi}}, \bibinfo {author} {\bibfnamefont {Z.}~\bibnamefont {Zhang}}, \bibinfo {author} {\bibfnamefont {L.}~\bibnamefont {You}}, \bibinfo {author} {\bibfnamefont {Y.}~\bibnamefont {Dong}}, \bibinfo {author} {\bibfnamefont {S.}~\bibnamefont {Das}}, \bibinfo {author} {\bibfnamefont {H.}~\bibnamefont {Zhang}}, \bibinfo {author} {\bibfnamefont {L.}~\bibnamefont {Zheng}}, \bibinfo {author} {\bibfnamefont {H.}~\bibnamefont {Wu}}, \bibinfo {author} {\bibfnamefont {W.}~\bibnamefont {Lv}}, \bibinfo {author} {\bibfnamefont {G.}~\bibnamefont {Xie}}, \bibinfo {author} {\bibfnamefont {X.}~\bibnamefont {Liu}}, \bibinfo {author} {\bibfnamefont
  {J.}~\bibnamefont {Li}}, \bibinfo {author} {\bibfnamefont {L.}~\bibnamefont {Chen}}, \bibinfo {author} {\bibfnamefont {L.-Q.}\ \bibnamefont {Chen}}, \bibinfo {author} {\bibfnamefont {W.}~\bibnamefont {Cao}},\ and\ \bibinfo {author} {\bibfnamefont {L.~W.}\ \bibnamefont {Martin}},\ }\bibfield  {title} {\enquote {\bibinfo {title} {{Mechanical-force-induced non-local collective ferroelastic switching in epitaxial lead-titanate thin films}},}\ }\href {https://doi.org/10.1038/s41467-019-11825-2} {\bibfield  {journal} {\bibinfo  {journal} {Nat. Commun.}\ } (\bibinfo {year} {2019}),\ 10.1038/s41467-019-11825-2}\BibitemShut {NoStop}%
\bibitem [{\citenamefont {Pandya}\ \emph {et~al.}(2019)\citenamefont {Pandya}, \citenamefont {Velarde}, \citenamefont {Gao}, \citenamefont {Everhardt}, \citenamefont {Wilbur}, \citenamefont {Xu}, \citenamefont {Maher}, \citenamefont {Agar}, \citenamefont {Dames},\ and\ \citenamefont {Martin}}]{Pandya2019-zb}%
  \BibitemOpen
  \bibfield  {author} {\bibinfo {author} {\bibfnamefont {S.}~\bibnamefont {Pandya}}, \bibinfo {author} {\bibfnamefont {G.~A.}\ \bibnamefont {Velarde}}, \bibinfo {author} {\bibfnamefont {R.}~\bibnamefont {Gao}}, \bibinfo {author} {\bibfnamefont {A.~S.}\ \bibnamefont {Everhardt}}, \bibinfo {author} {\bibfnamefont {J.~D.}\ \bibnamefont {Wilbur}}, \bibinfo {author} {\bibfnamefont {R.}~\bibnamefont {Xu}}, \bibinfo {author} {\bibfnamefont {J.~T.}\ \bibnamefont {Maher}}, \bibinfo {author} {\bibfnamefont {J.~C.}\ \bibnamefont {Agar}}, \bibinfo {author} {\bibfnamefont {C.}~\bibnamefont {Dames}},\ and\ \bibinfo {author} {\bibfnamefont {L.~W.}\ \bibnamefont {Martin}},\ }\bibfield  {title} {\enquote {\bibinfo {title} {{Understanding the role of ferroelastic domains on the pyroelectric and electrocaloric effects in ferroelectric thin films}},}\ }\href {https://doi.org/10.1002/adma.201803312} {\bibfield  {journal} {\bibinfo  {journal} {Adv. Mater.}\ } (\bibinfo {year} {2019}),\ 10.1002/adma.201803312}\BibitemShut {NoStop}%
\bibitem [{\citenamefont {Pompe}\ \emph {et~al.}(1993)\citenamefont {Pompe}, \citenamefont {Gong}, \citenamefont {Suo},\ and\ \citenamefont {Speck}}]{Pompe1993-dp}%
  \BibitemOpen
  \bibfield  {author} {\bibinfo {author} {\bibfnamefont {W.}~\bibnamefont {Pompe}}, \bibinfo {author} {\bibfnamefont {X.}~\bibnamefont {Gong}}, \bibinfo {author} {\bibfnamefont {Z.}~\bibnamefont {Suo}},\ and\ \bibinfo {author} {\bibfnamefont {J.~S.}\ \bibnamefont {Speck}},\ }\bibfield  {title} {\enquote {\bibinfo {title} {{Elastic energy release due to domain formation in the strained epitaxy of ferroelectric and ferroelastic films}},}\ }\href {https://doi.org/10.1063/1.355215} {\bibfield  {journal} {\bibinfo  {journal} {J. Appl. Phys.}\ } (\bibinfo {year} {1993}),\ 10.1063/1.355215}\BibitemShut {NoStop}%
\bibitem [{\citenamefont {Ehara}\ \emph {et~al.}(2017)\citenamefont {Ehara}, \citenamefont {Yasui}, \citenamefont {Oikawa}, \citenamefont {Shiraishi}, \citenamefont {Shimizu}, \citenamefont {Tanaka}, \citenamefont {Kanenko}, \citenamefont {Maran}, \citenamefont {Yamada}, \citenamefont {Imai}, \citenamefont {Sakata}, \citenamefont {Valanoor},\ and\ \citenamefont {Funakubo}}]{Ehara2017-cs}%
  \BibitemOpen
  \bibfield  {author} {\bibinfo {author} {\bibfnamefont {Y.}~\bibnamefont {Ehara}}, \bibinfo {author} {\bibfnamefont {S.}~\bibnamefont {Yasui}}, \bibinfo {author} {\bibfnamefont {T.}~\bibnamefont {Oikawa}}, \bibinfo {author} {\bibfnamefont {T.}~\bibnamefont {Shiraishi}}, \bibinfo {author} {\bibfnamefont {T.}~\bibnamefont {Shimizu}}, \bibinfo {author} {\bibfnamefont {H.}~\bibnamefont {Tanaka}}, \bibinfo {author} {\bibfnamefont {N.}~\bibnamefont {Kanenko}}, \bibinfo {author} {\bibfnamefont {R.}~\bibnamefont {Maran}}, \bibinfo {author} {\bibfnamefont {T.}~\bibnamefont {Yamada}}, \bibinfo {author} {\bibfnamefont {Y.}~\bibnamefont {Imai}}, \bibinfo {author} {\bibfnamefont {O.}~\bibnamefont {Sakata}}, \bibinfo {author} {\bibfnamefont {N.}~\bibnamefont {Valanoor}},\ and\ \bibinfo {author} {\bibfnamefont {H.}~\bibnamefont {Funakubo}},\ }\bibfield  {title} {\enquote {\bibinfo {title} {{In-situ observation of ultrafast 90\textdegree{} domain switching under application of an electric field in (100)/(001)-oriented
  tetragonal epitaxial Pb(Zr0.4Ti0.6)O3 thin films}},}\ }\href {https://doi.org/10.1038/s41598-017-09389-6} {\bibfield  {journal} {\bibinfo  {journal} {Sci. Rep.}\ } (\bibinfo {year} {2017}),\ 10.1038/s41598-017-09389-6}\BibitemShut {NoStop}%
\bibitem [{\citenamefont {Zhang}\ \emph {et~al.}(2019)\citenamefont {Zhang}, \citenamefont {Han}, \citenamefont {Garlow}, \citenamefont {Tan}, \citenamefont {Xue}, \citenamefont {Chen}, \citenamefont {Munroe}, \citenamefont {Valanoor},\ and\ \citenamefont {Zhu}}]{Zhang2019-dl}%
  \BibitemOpen
  \bibfield  {author} {\bibinfo {author} {\bibfnamefont {Y.}~\bibnamefont {Zhang}}, \bibinfo {author} {\bibfnamefont {M.-G.}\ \bibnamefont {Han}}, \bibinfo {author} {\bibfnamefont {J.~A.}\ \bibnamefont {Garlow}}, \bibinfo {author} {\bibfnamefont {Y.}~\bibnamefont {Tan}}, \bibinfo {author} {\bibfnamefont {F.}~\bibnamefont {Xue}}, \bibinfo {author} {\bibfnamefont {L.-Q.}\ \bibnamefont {Chen}}, \bibinfo {author} {\bibfnamefont {P.}~\bibnamefont {Munroe}}, \bibinfo {author} {\bibfnamefont {N.}~\bibnamefont {Valanoor}},\ and\ \bibinfo {author} {\bibfnamefont {Y.}~\bibnamefont {Zhu}},\ }\bibfield  {title} {\enquote {\bibinfo {title} {{Deterministic ferroelastic domain switching using ferroelectric bilayers}},}\ }\href {https://doi.org/10.1021/acs.nanolett.9b01782} {\bibfield  {journal} {\bibinfo  {journal} {Nano Lett.}\ } (\bibinfo {year} {2019}),\ 10.1021/acs.nanolett.9b01782}\BibitemShut {NoStop}%
\end{thebibliography}%
